\documentclass[sn-nature]{sn-jnl}
\usepackage{graphicx}%

\usepackage{multirow}%
\usepackage{amsmath,amssymb,amsfonts}%
\usepackage{amsthm}%
\usepackage{mathrsfs}%
\usepackage[title]{appendix}%
\usepackage{xcolor}%
\usepackage{textcomp}%
\usepackage{manyfoot}%
\usepackage{booktabs}%
\usepackage[separate-uncertainty=true]{siunitx}
\usepackage{caption}
\usepackage{subcaption}
\usepackage{algorithm}
\usepackage{algpseudocode}
\usepackage{algorithmicx}%
\usepackage{makecell}
\usepackage{array}
\usepackage{needspace}      

\usepackage[normalem]{ulem}

\usepackage{xr}
\begin{document}

\title[Article Title]{Virtual ultrasound machine operating in a GHz to MHz frequency range for particle-based biomedical simulations}

\author[1,2]{\fnm{Urban} \sur{\v{C}oko}}\email{urban.coko@ki.si}

\author[1,2]{\fnm{Tilen} \sur{Potisk}}\email{tilen.potisk@ki.si}

\author*[1,2,3]{\fnm{Matej} \sur{Praprotnik}}\email{praprot@cmm.ki.si}

\affil[1]{\orgdiv{Theory Department}, \orgname{National Institute of Chemistry}, \orgaddress{\street{Hajdrihova 19}, \postcode{SI-1001} \city{Ljubljana}, \country{Slovenia}}}

\affil[2]{\orgdiv{Department of Physics}, \orgname{Faculty of Mathematics and Physics, University of Ljubljana}, \orgaddress{\street{Jadranska 19}, \postcode{SI-1000} \city{Ljubljana}, \country{Slovenia}}}

\affil[3]{\orgdiv{Universitat de Barcelona Institute of Complex Systems (UBICS)},
\newline
\orgaddress{\street{C/ Martí i Franqués 1}, \postcode{08028} \city{Barcelona}, \country{Spain}}}

\abstract{Ultrasound–matter interactions underpin numerous biomedical and soft-matter applications, yet simulating these phenomena is challenging due to the large separation of viscous and sonic time scales. Continuum methods capture large-scale wave propagation but cannot resolve microscale interactions, while particle-based approaches offer molecular resolution but struggle with efficiency and stability at larger scales. We introduce a particle-based virtual ultrasound machine that uses a novel smoothed dissipative particle dynamics variant with an implicit pressure solver and a negative-pressure stabilization scheme, required to mimic acoustic propagation across MHz–GHz frequencies. We demonstrate its capabilities by modeling the acoustophoresis of encapsulated microbubbles, a key mechanism in ultrasound-mediated drug delivery. Beyond this application, the approach establishes a generalizable platform for simulating wave–matter interactions in soft and biological materials, opening new directions for computational studies of acoustics-driven phenomena in science and engineering.
}


\maketitle

\section{Introduction}
\noindent

In modern medicine, ultrasound (US) has become an indispensable modality, used both for imaging and therapeutic purposes. For imaging applications, its most advantageous features are the absence of ionizing radiation and relatively low purchase and maintenance costs \cite{Moran2020}. It is applied extensively in soft-tissue imaging—from fetal monitoring to echocardiography \cite{Moran2020}—and in therapeutic settings, where focused US enables tumor ablation and low-intensity US promotes tissue healing \cite{Best2016}.

The frequency ranges of medical US vary with application. Therapeutic US generally uses lower frequencies (typically 0.5–3 MHz) \cite{Miller2012}. Diagnostic US typically operates between \SI{2}{MHz} and \SI{40}{MHz} \cite{Neumann2018}, with lower frequencies (1–5 MHz) providing greater penetration depth (e.g. abdominal or cardiac scans), and higher frequencies (20–40 MHz) offering better spatial resolution for superficial structures (e.g. skin or small animal imaging) \cite{Neumann2018, Moran2020}. For instance, at \qtyrange{15}{20}{MHz}, the imaging depth is about \qtyrange{3}{4}{cm}, whereas at \SI{50}{MHz} it decreases to \SI{9}{mm} \cite{Moran2020,Maresca2018a}. Acoustic excitations in the 100 MHz–GHz range are largely limited to research applications. GHz US can probe subcellular and molecular-scale motions; for example, GHz acoustic pulses have been shown to stimulate neuronal ion channels \cite{Balasubramanian2020}. Moreover, protein vibrational modes often lie in the GHz–THz range, with some low-frequency modes strongly correlated with large-amplitude conformational changes triggered by ligand binding \cite{Tama2001, Tobi2005, Wako2011}. 

For diagnostic imaging, contrast agents must exhibit resonance within the diagnostic frequency range. Among the most widely used agents are encapsulated microbubbles (EMBs)—gas-filled spheres encapsulated by lipid or protein shells, typically a few micrometers in diameter \cite{Navarro2023, Yusefi2022}. When exposed to US, EMBs undergo oscillations that produce strong, distinctive acoustic signals, enabling clear contrast between blood flow and surrounding tissues \cite{Yusefi2022,Heiles2021,Maresca2018a}. Beyond imaging, EMBs can serve as vehicles for targeted drug delivery: under sufficiently high acoustic pressures, their shells rupture through cavitation, inducing localized mechanical stress that temporarily increases cell membrane permeability—a process known as sonoporation \cite{Navarro2023}. This phenomenon facilitates site-specific delivery of therapeutic agents, reducing systemic dosage and minimizing side effects. Furthermore, EMBs can be functionalized with ligands that bind to disease-specific biomarkers, allowing them to accumulate selectively at pathological sites such as tumor vasculature \cite{Navarro2023}.

Conducting experimental studies in this domain is far from trivial. The parameters of EMBs are typically constrained to those achievable in laboratory conditions, and generating US itself requires specialized equipment. Therapeutic US systems generally operate at a single frequency or a limited set of discrete frequencies and deliver average intensities up to \SI{3}{W/cm^2} \cite{TerHaar1999}. While intensity can be adjusted either discretely or continuously, the range of accessible intensities in practice remains narrow. Consequently, exploring a broad parameter space experimentally is challenging. To overcome these limitations, numerical simulations are frequently employed as a preliminary platform for investigation, effectively serving as a virtual laboratory before physical experiments are performed.

Designing a virtual US machine requires a fluid model that accurately captures both the viscosity and the acoustic properties of water at micrometer scale. This is difficult in practice because, as the system size increases, viscous and acoustic timescales become increasingly separated, rendering the problem multi-scale (see Section~\ref{sec:sdpd}). Traditionally, such scenarios have been tackled using hybrid frameworks—such as PROTEUS \cite{Heiles2025}—that employ distinct solvers for sound propagation and fluid dynamics, coupling them through interactions with immersed structures. Common approaches for acoustic wave modeling include k-Wave \cite{Treeby2010}, Field~II \cite{Jensen1997}, and SIMUS \cite{Garcia2022, Cigier2022, Garcia2024}, while fluid flow is typically handled using continuum, mesh-based methods such as the lattice-Boltzmann method (LBM) and various computational fluid dynamics (CFD) techniques.
In some cases, numerical methods are coupled with analytical solutions to reduce computational demands \cite{Pavlic2025}. While hybrid approaches are useful for certain special use-cases, their implementation is highly complex, system-dependent, and employs numerous simplifications. 

On the other hand, particle-based fluid descriptions offer a compelling alternative to mesh-based methods, especially for biophysical simulations. Most notably, they are inherently compatible with particle-based models of immersed structures, since both are Lagrangian in nature. In contrast, coupling Lagrangian particle-based structures with Eulerian mesh-based fluids often necessitates complex interpolation schemes, such as Peskin’s Immersed Boundary Method \cite{Peskin2002}, which can compromise both model accuracy and simplicity. Furthermore, mesh-based methods generally struggle to resolve fine structural details unless the grid spacing is significantly smaller than the features of interest or adaptive meshing is employed. Particle-based methods also often provide a straigthforward implementation of multiphase systems and thermal fluctuations, which are more challenging to implement in mesh-based formulations. Finally, dense suspensions are typically represented with higher fidelity in particle-based simulations, further underscoring their suitability for complex biophysical environments.
Interestingly, particle-based US simulations and experimental approaches exhibit fundamentally opposing challenges. In experimental settings, achieving lower frequencies in the MHz range is relatively straightforward, whereas generating THz frequencies poses significant problems. The inverse holds for particle-based simulation approaches. Particle-based simulations in lower-frequency regimes often suffer from numerical freezing artifacts \cite{Trofimov2003}, whereas experimental implementations frequently encounter issues related to unwanted sample heating \cite{Cheng2019}.

\begin{figure}[h!]
    \centering
    \includegraphics[width=0.9\textwidth]{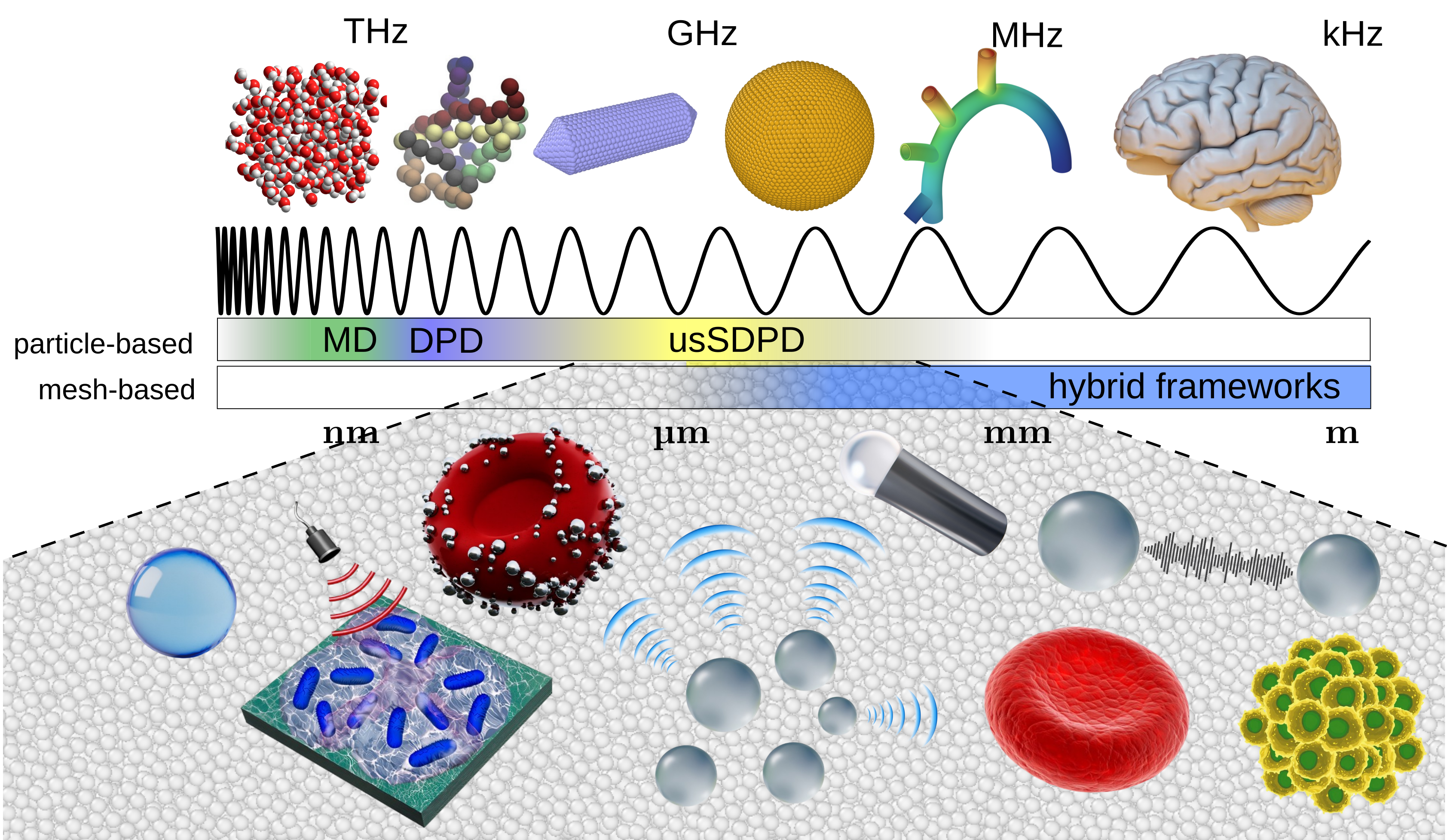}
    \caption{\textbf{Overview of spatiotemporal scales, simulation methods and systems, studied with US.} Top: commonly used methods for US simulations and the length- and frequency-scales they typically describe. Boundaries between methods are approximate, as the applicable range depends on computational resources, geometry, and optimizations. Representative systems are shown below the frequencies: molecular dynamics (MD) for all-atom simulations of water, DPD for THz excitations in proteins, and hybrid frameworks for large-scale biological tissue simulations (such as capillary flow) and EMB dynamics. Our method, usSDPD, fits on the medically-relevant scale of EMBs. This figure shows the ranges, where the viscosity and compressibility match those of water. For example, the DPD method might be used also at larger scales, but either the compressibility and viscosity of water would not be correct, or a liquid with lower compressibility and/or higher viscosity would be used.
    Bottom: schematic representations of structures that can be coupled with usSDPD, including (from left to right) hydrogel microspheres, biofilms \cite{Huang2025}, red blood cell microrobots with magnetic nanoparticles \cite{Huang2023}, scattering EMBs, tubular microrobots \cite{Lu2020}, red blood cells, interacting oscillating EMBs, tumor spheroids, and others.}
    \label{fig:overview}
\end{figure}

In this paper, we design a particle-based virtual US machine capable of simulating US propagation at the micrometer scale (MHz frequencies), thereby enabling more versatile \textit{in silico} experimentation. We achieve this by developing a new SDPD-based method called usSDPD, which is the first particle-based method that can faithfully capture both the viscosity and the speed of sound of water at micrometer scales, while also minimizing elastic artifacts that typically appear under highly negative pressures. This method makes it possible to study synthetic and biological systems previously inaccessible to particle-based approaches (Figure \ref{fig:overview}). It enables the modeling of dense suspensions of immersed structures and their interactions without introducing excessive technical complexity.
The virtual US machine presented here is envisioned as a core component of a general virtual US framework. This framework will integrate the virtual US machine with detailed models of immersed structures, such as EMBs and gas vesicles \cite{Ntarakas2025}, enabling comprehensive simulations of US interactions with complex biological and synthetic systems. Together, these elements will provide a versatile platform for exploring US-based imaging, therapy, and mechanistic studies entirely in silico.

\needspace{5\baselineskip}
\section{Results and discussion}
\label{sec: Results}
\noindent

We present a particle-based fluid simulation method, termed usSDPD, that enables accurate modeling of US wave propagation in weakly compressible fluids such as water. On this basis, we construct a virtual US machine to simulate a sample immersed in water and positioned between two opposing US transducers. The virtual US machine is applied to simulate EMB acoustophoresis, demonstrating its ability to capture US-driven particle dynamics and illustrating its potential as a versatile tool for investigating a broad range of US-mediated phenomena.

\subsection{Fluid simulation}
\label{sec:fluid-simulation}

We assess the usSDPD method by characterizing its stability, viscosity, and pressure behavior at granularities of \SI{0.01}{\um}, \SI{0.1}{\um}, and \SI{1}{\um}, where granularity denotes the fluid particle diameter. Appropriate timescales for each granularity are provided in the Supplementary Information (Section \ref{sec:scaling-of-dimensionless-quantities}), along with detailed analyses and parameter tables (Sections \ref{sec:analysis-of-the-fluid-in-equilibrium} and \ref{sec:parameters}). Our results indicate that a granularity of \SI{1}{\um} and higher is not sufficiently accurate for our intended application. 
At \SI{0.1}{\um} granularity, our method exhibits a substantial gain in numerical stability, enabling a 40-fold increase in the maximum stable timestep, while inducing only a minor rise in viscosity. In contrast, simulations at \SI{0.01}{\um} do not exhibit comparable improvements in timestep length, and further refinement to even smaller granularities is not advantageous, as DPD-based methods are inherently more suitable at nanoscopic scales. Consequently, usSDPD is best applied at granularities of \SI{0.01}{\um} and \SI{0.1}{\um}, with the greatest benefits observed at \SI{0.1}{\um} granularity.

\begin{figure}[h!]
    \centering
    \begin{subfigure}[t]{0.8\textwidth}
        \centering
        \caption{}
        \includegraphics[width=\linewidth]{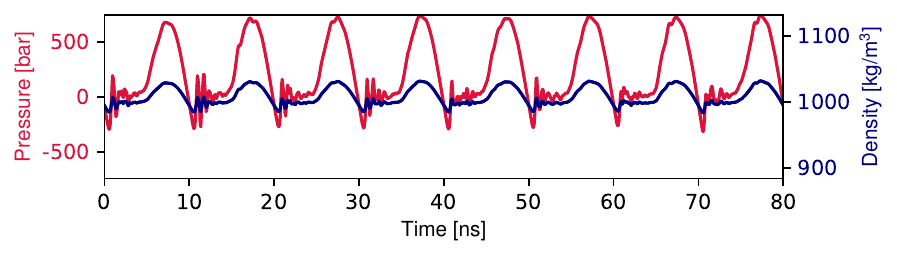}
        \label{fig:denpres-wave-unstable}
    \end{subfigure}%

    \vspace{-2.5em}

    \begin{subfigure}[t]{0.8\textwidth}
        \centering
        \caption{}
        \includegraphics[width=\linewidth]{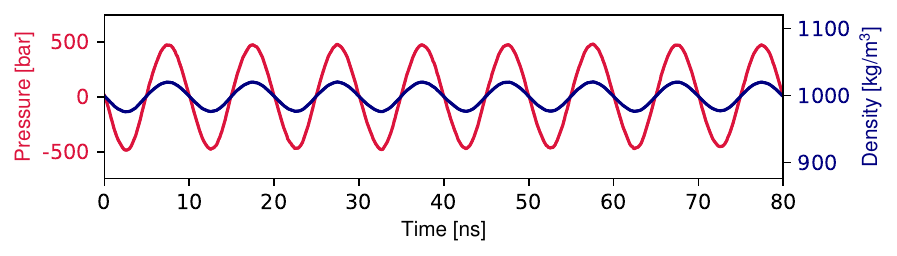}
        \label{fig:denpres-wave-stable}
    \end{subfigure}%

    \caption{\textbf{Response of (a) standard (vanilla) SDPD and (b) usSDPD fluids to bulk volume oscillation.} Both simulations are done at \SI{0.1}{\um} granularity with both speed of sound and viscosity matched to those of water. usSDPD shows more stable behavior than vanilla SDPD at negative pressures. The pressure amplitude is large compared to the density amplitude, which is characteristic of water with low compressibility. In vanilla SDPD simulations, 10 times smaller timestep is used for stability.}
    \label{fig:denpres-wave}
\end{figure}

Next, we demonstrate that although the SDPD formulation exhibits significant deficiencies when applied to the virtual US machine, the proposed usSDPD approach remains robust. To assess its performance, we simulate a fluid block enclosed by periodic boundary conditions (PBCs) in all directions, where the simulation cell volume is periodically oscillated to generate bulk pressure oscillations. These volumetric oscillations induce corresponding variations in pressure and density, enabling direct assessment of the method’s response to large-amplitude acoustic forcing.
As shown in Figure \ref{fig:denpres-wave}, the SDPD formulation becomes unstable at negative pressures, whereas the usSDPD method remains stable over a broad range spanning both positive and negative pressures throughout the entire oscillatory cycle. This indicates that the usSDPD method can successfully sustain strong periodic compressions and rarefactions without breakdown.

\subsection{Virtual ultrasound machine}
\label{sec:virtual-ultrasound-machine}

\begin{figure}[h!]
    \centering
    \includegraphics[width=0.8\textwidth]{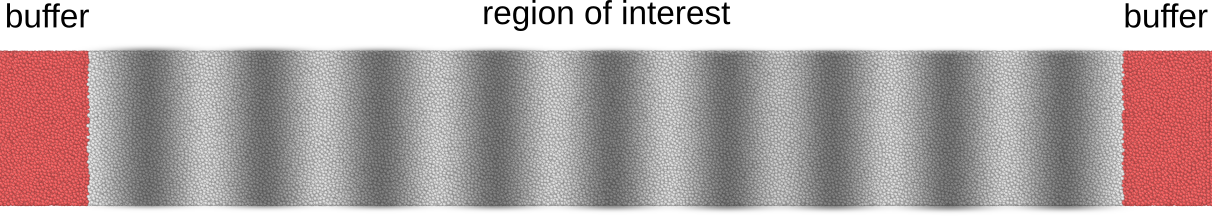}
    \caption{\textbf{Schematic of a virtual US machine.} The machine consists of two US sources positioned at the left and right boundaries of the open simulation box (buffers), as well as the region of interest, where the immersed structure is located. By imposing oscillations of the same frequency in the left and right buffers, a standing wave field is generated in the region of interest (dark and light regions correspond to low and high density, respectively).}
    \label{fig:obmd}
\end{figure}

A particle-based virtual US machine consists of two key components: a particle-based fluid simulation method capable of reproducing fluid compressibility at the target length scale, and a mechanism for imposing pressure oscillations at the boundaries of the simulation domain, representing a simplified model of a US transducer. For the fluid dynamics at mesoscopic granularities (\SI{0.01}{\um}, \SI{0.1}{\um}) and low-compressibility fluids such as water (MHz to GHz frequencies), we employ the previously introduced usSDPD method. At nanoscopic length scales (THz frequencies), however, DPD has been shown to be more appropriate; this is discussed in detail in Supplementary Information, Section~\ref{sec:analysis-of-the-fluid-in-equilibrium}, and such virtual US machines may therefore employ DPD as a fluid simulation method.
The virtual US machine considered here is realized by placing two transducers on opposing sides of the simulation box, which confines both the fluid and the immersed structure in one direction, while enforcing PBCs at the remaining two spatial directions. The transducers are modeled using the open boundary molecular dynamics (OBMD) method (Section~\ref{sec:OBMD}), which represents an open system of particles \cite{Buscalioni2015} (Fig.~\ref{fig:obmd}). In addition to generating pressure oscillations, OBMD enables the application of controlled shear stresses, a capability not available in conventional US transducers and particularly useful for investigating shear-driven flows.

\subsection{Acoustophoresis simulation}
Our virtual US machine is designed to simulate US–fluid–structure interactions in weakly compressible fluids, such as water from nanometer to micrometer granularities. As a representative application, we study the dynamics of a EMB in a standing US field, where its motion is driven by acoustic radiation forces. These forces originate from the mismatch in compressibility and density between the EMB and the surrounding fluid and give rise to acoustophoresis, i.e., migration of the EMB within the acoustic field \cite{Gorkov1962}. In a standing wave, acoustic radiation forces drive the migration of immersed structures toward either pressure nodes or pressure antinodes, the latter typical for EMBs and the former typical for immersed beads. This phenomenon, known as acoustophoresis, enables precise acoustic manipulation and is widely used in microfluidics for particle sorting, purification \cite{Shields2015}, concentration \cite{Jakobsson2015}, washing \cite{Deshmukh2014}, enrichment \cite{Magnusson2017}, and confinement trapping \cite{Ohlsson2016}. It is also the driving mechanism behind acoustic tweezers \cite{Shi2009}. Analytical models \cite{King1934, Yosioka1955, Gorkov1962, Leibacher2014} and numerical simulations \cite{Sepehrirahnama2015, Haydock2005}, including finite element methods \cite{Wang2009}, have long been employed to study these forces.

\begin{figure}[h!]
    \centering
    \includegraphics[width=0.6\textwidth]{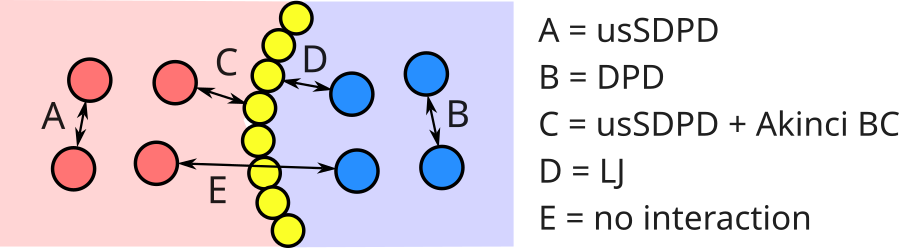}
    \caption{\textbf{Interactions in the acoustophoresis simulation system.} Red circles represent external fluid particles, yellow circles represent membrane particles, and blue circles represent internal fluid particles. The arrows denote interactions between particle types.}
    \label{fig:interactions}
\end{figure}

The model of the experimental system includes the external fluid, the EMB membrane, and the internal fluid within the EMB, see Figure \ref{fig:interactions}. For the external fluid we employ the usSDPD method, which allows pressure waves of large amplitude to be simulated without introducing excessive spurious elasticity, Section \ref{sec:fluid-simulation}. The membrane is described using an isotropic elastic model \cite{Ntarakas2025, Vlachomitrou2017}, which places membrane particles in the vertices of the triangulated EMB surface, and calculates forces by taking derivatives of the total elastic energy with respect to vertex positions. For a detailed list of parameters, see Supplementary Information, Section \ref{sec:parameters}. Most notably, in our setup, the EMB is modeled as neutrally buoyant to allow for a larger timestep. However, its internal fluid remains significantly more compressible than water. Because of this high compressibility, we adopt the standard DPD method rather than SDPD for modeling the internal fluid. This reduces both the computational overhead and the number of parameters in the simulation. Moreover, it highlights the modularity of the particle-based framework, wherein distinct fluid models can be seamlessly coupled. In contrast, integrating a particle-based EMB model with a mesh-based fluid solver would require the use of interpolation techniques, such as Peskin’s Immersed Boundary Method \cite{Peskin2002}, to mediate interactions across the particle-mesh interface. Such approaches introduce additional algorithmic complexity and may compromise numerical accuracy.
The final component of the model involves specifying the interactions between different types of particles. We assume an impermeable EMB membrane. For interactions between the external fluid and the membrane, we use the Akinci boundary condition \cite{Akinci2013}, commonly employed in SPH for incompressible and weakly compressible fluids. This boundary condition has the important property that the boundary force is invariant with respect to the sampling density (resolution) of the boundary, which is crucial for an oscillating EMB whose sampling density changes dynamically. It should also be noted that Monaghan’s artificial pressure lowers the pressure only for non-boundary fluid particles; Akinci boundary particles still experience the full pressure force. Because ordinary DPD differs fundamentally from SDPD, the same boundary condition cannot be applied to interactions between the internal fluid and membrane particles. In this case, we employ a simple Lennard-Jones interaction (Figure \ref{fig:interactions}).

\begin{figure}[H]

    \begin{minipage}[t]{0.6\textwidth}
        \centering
        \begin{subfigure}[t]{\textwidth}
            \caption{}
            \includegraphics[width=\textwidth]{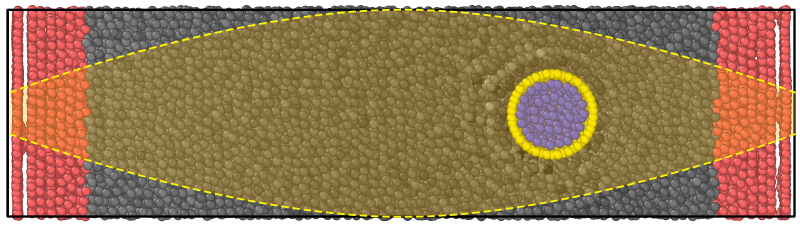}
            \label{fig:acoustophoresis-snapshot-beginning}
        \end{subfigure}%

        \vspace{-1.5em}

        \begin{subfigure}[t]{\textwidth}
            \caption{}
            \includegraphics[width=\textwidth]{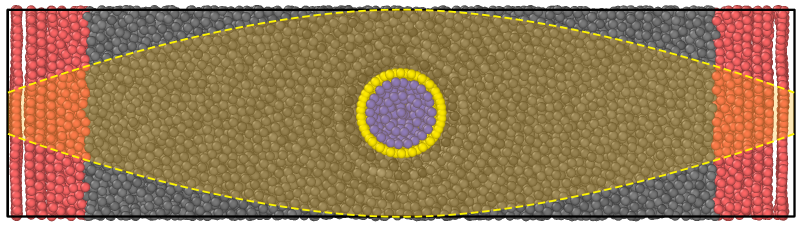}
            \label{fig:acoustophoresis-snapshot-end}
        \end{subfigure}
    \end{minipage}
    \hfill
    \begin{minipage}[t]{0.4\textwidth}
        \centering
        \begin{subfigure}[t]{\textwidth}
            \caption{}
            \vspace{-.5em}
            \includegraphics[width=\textwidth]{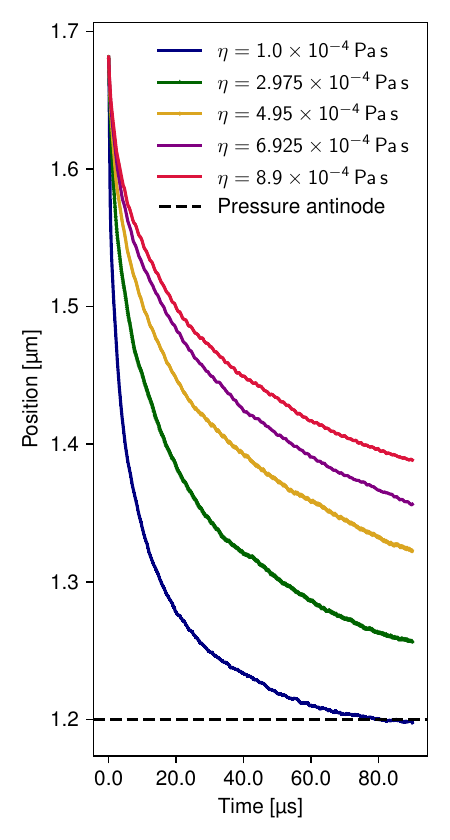}
            \label{fig:acoustophoresis-trajectory}
        \end{subfigure}
    \end{minipage}
    
    \vspace{-13.8em}

    \begin{minipage}[t]{0.6\textwidth}
        \begin{subfigure}[t]{0.53\textwidth}
            \caption{}
            \vspace{-.7em}
            \includegraphics[width=\textwidth]{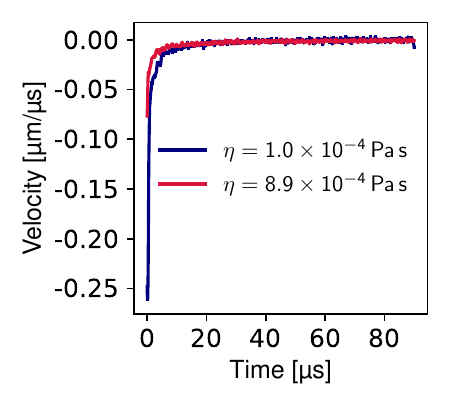}
            \label{fig:acoustophoresis-velocity}
        \end{subfigure}%
        \hfill
        \begin{subfigure}[t]{0.46\textwidth}
            \caption{}
            \vspace{-1.5em}
            \includegraphics[width=\textwidth]{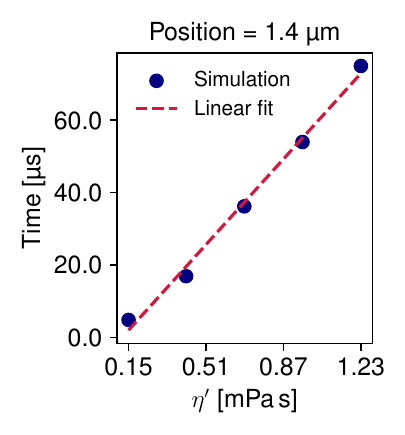}
            \label{fig:acoustophoresis-viscosity-bruus-fit}
        \end{subfigure}
    \end{minipage}

    \caption{\textbf{Acoustophoresis simulation in a $\SI{2.4}{\um} \times \SI{0.63}{\um} \times \SI{0.63}{\um}$  cell.} 
    (a) snapshot of the acoustophoresis simulation at the beginning and 
    (b) the end of the simulation (the system is sliced in half along the page plane to improve visibility). Buffer regions are shown in red and the yellow shading schematically illustrates the pressure field of the standing wave.
    (c) Trajectories of the EMB center-of-mass for input viscosities ranging uniformly from $\eta = \SI{8.9e-4}{\pascal \second}$ to $\eta = \SI{1.0e-4}{\pascal \second}$ with pressure amplitude $\Delta p_0 = \SI{9}{bar}$ in all cases. Frequency is \SI{178}{MHz} and EMB diameter is \SI{0.24}{\um} in all cases. 
    (d) EMB center-of-mass velocity for largest and smallest viscosity. 
    (e) Time required for a EMB to reach the position $x = \SI{1.4}{\um}$ as a function of the simulated viscosity $\eta'$. As explained in the main text, simulated viscosity differs slightly from the input viscosity due to additional LJ contribution.}
\label{fig:acoustophoresis}

\end{figure}

Finally, we apply this modeling framework to simulate acoustophoresis in a standing wave for an open system of length \SI{2.4}{\um} (see Figure \ref{fig:acoustophoresis}). The standing wave is generated by enforcing in-phase harmonic oscillations in the buffer regions. Simulations are performed at five uniformly spaced viscosities ranging from $\eta = \SI{1.0e-4}{\pascal \second}$ to $\eta = \SI{8.9e-4}{\pascal \second}$. In all cases, the EMB tends toward the pressure antinode; however, the characteristic migration time is shorter at lower viscosities. According to the theoretical model presented in \cite{Bruus2012}, the time required for a compressible particle to traverse a given distance scales linearly with viscosity—a trend we also observe in the simulation (see Figure \ref{fig:acoustophoresis-viscosity-bruus-fit}).
In these simulations, fewer than four iterations of the implicit-compressible pressure solver are typically required, indicating rapid pressure convergence. In general, the number of required iterations increases with increasing positive or negative pressures, but empirically we find it to remain below ten for all practical cases. Total system momentum is conserved in the simulation, which means that the external fluid acquires a net velocity as the EMB moves toward the pressure antinode. To prevent drift of the external fluid, we enforce zero net momentum by subtracting the center-of-mass (COM) velocity of the external fluid particles at every timestep.

\begin{figure}[H]
    \begin{subfigure}[t]{\textwidth}
        \centering
        \caption{}
        \includegraphics[width=\textwidth]{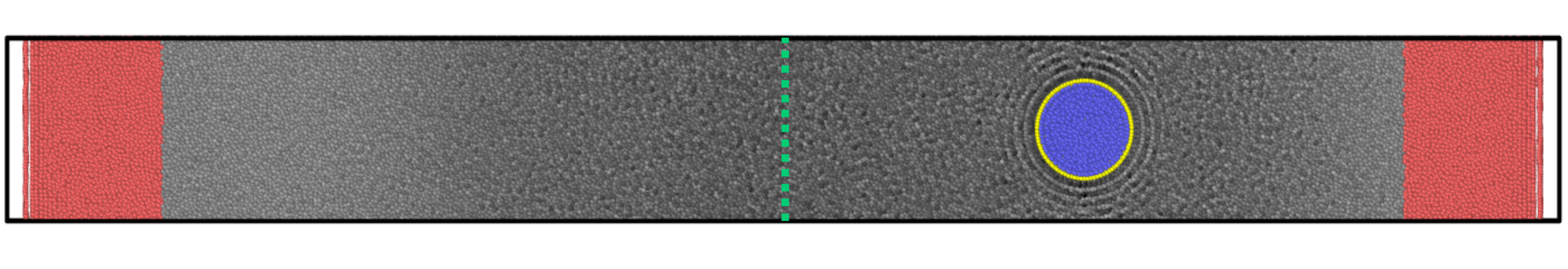}
        \label{fig:acoustophoresis-large-snapshot-beginning}
    \end{subfigure}

    \vspace{-1.5em}

    \begin{subfigure}[t]{\textwidth}
        \centering
        \caption{}
        \includegraphics[width=\textwidth]{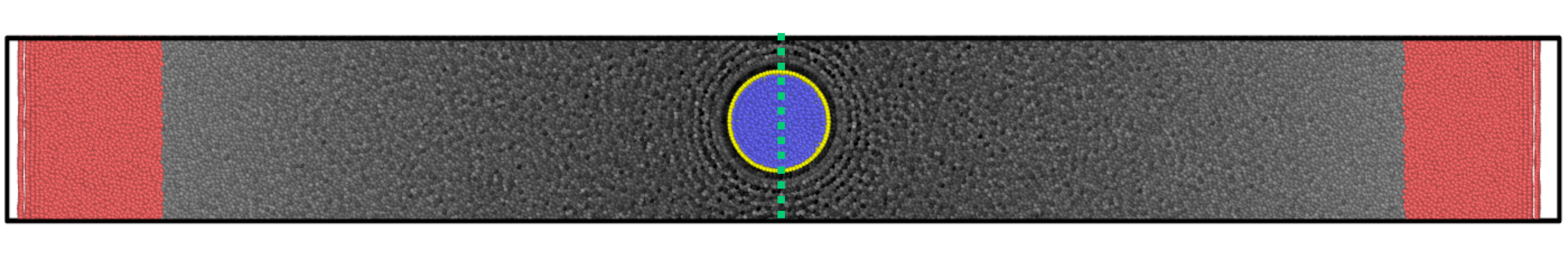}
        \label{fig:acoustophoresis-large-snapshot-end}
    \end{subfigure}

    \vspace{-1.7em}

    \begin{minipage}[t]{0.55\textwidth}
        \centering
        \begin{subfigure}[t]{\textwidth}
            \caption{}
            \includegraphics[width=\textwidth]{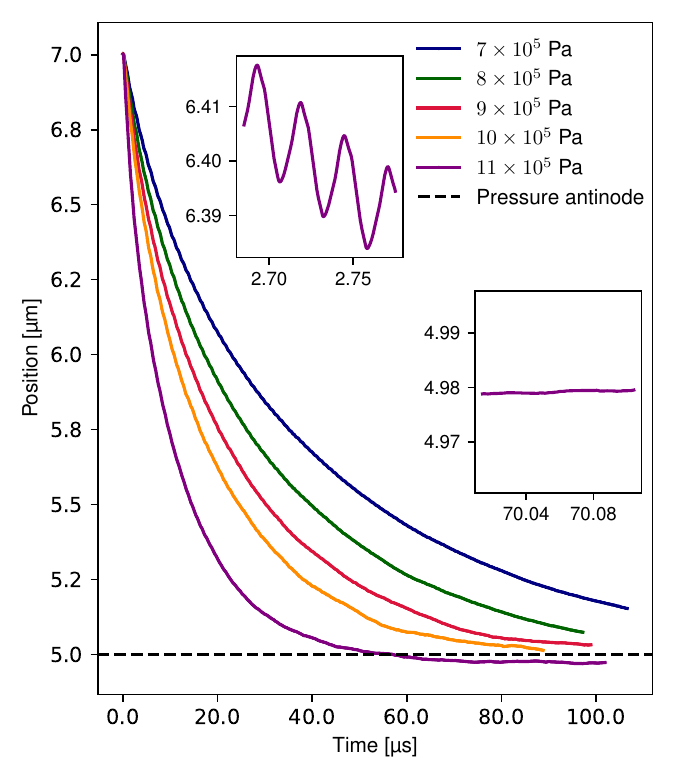}
            \label{fig:acoustophoresis-large-trajectories}
        \end{subfigure}
    \end{minipage}
    \hfill
    \begin{minipage}[t]{0.44\textwidth}
        \centering
        \begin{subfigure}[t]{\textwidth}
            \caption{}
            \includegraphics[width=\textwidth]{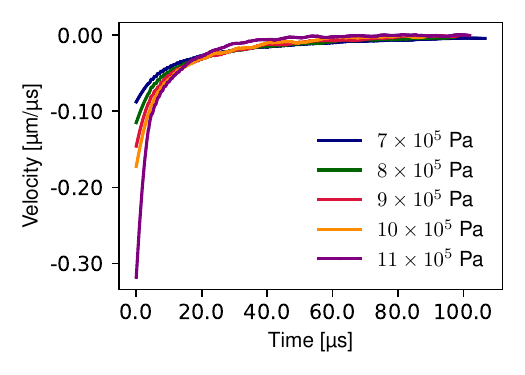}
            \label{fig:acoustophoresis-large-velocities}
        \end{subfigure}%

        \vspace{-3.5em}

        \begin{subfigure}[t]{\textwidth}
            \caption{}
            \includegraphics[width=\textwidth]{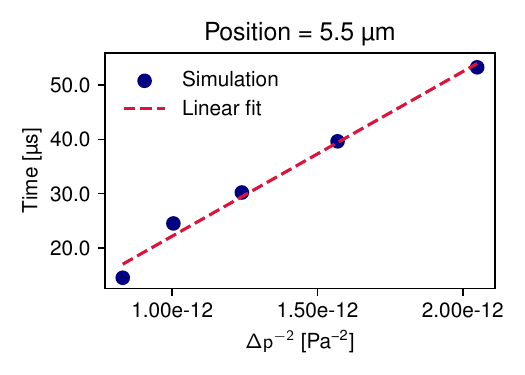}
            \label{fig:acoustophoresis-bruus-fit}
        \end{subfigure}
    \end{minipage}

    \caption{\textbf{Acoustophoresis simulations in a $\SI{10}{\um} \times \SI{1.2}{\um} \times \SI{1.2}{\um}$ cell.} 
    (a) snapshot of the acoustophoresis simulation at the beginning and (b) the end of the simulation (the simulation system is sliced in half along the page plane to increase visibility). Buffer regions are shown in red and the green dashed line denotes the pressure antinode. 
    (c) Trajectories of the EMB center-of-mass with input viscosity $\eta = \SI{1.0e-4}{\pascal \second}$, frequency \SI{38.6}{MHz} and EMB diameter \SI{0.6}{\um}. Pressure amplitudes are given in the figure. Insets show a zoomed-in part of the curve with the highest pressure amplitude. 
    (d) EMB center-of-mass velocity, corresponding to the position plot in (c). 
    (e) Time required for a EMB to reach the position $x = \SI{5.7}{\um}$ as a function of the inverse square of the pressure amplitude.}
    \label{fig:acoustophoresis-large}
\end{figure}

To demonstrate the scalability of our method, we perform a larger simulation at a frequency of \SI{39}{MHz}, which falls within the medical US regime \cite{Neumann2018}, see Figure~\ref{fig:acoustophoresis-large}. In this case, the simulation cell length is increased to \SI{10}{\um}, the width to \SI{1.2}{\um}, and the EMB diameter to \SI{0.6}{\um}. The pressure amplitude is gradually ramped up to its final value over the first \SI{0.15}{\us}. We observe that the EMB consistently settles at the pressure antinode. To investigate the dynamics, we compare EMB trajectories under three different pressure amplitudes. According to \cite{Bruus2012}, the time required for a compressible particle to traverse a given distance scales as $\Delta p^{-2}$. While our simulations generally follow this trend, we observe a slight deviation from the expected $\Delta p^{-2}$ scaling at higher pressure amplitudes (\SI{10}{bar} and \SI{11}{bar}), suggesting that additional factors may influence EMB migration under these conditions.
We also observe that the COM motion of the EMB shifts from oscillatory pattern at short times (when EMB velocity is high) towards a constant value at longer times (when EMB velocity is low), see insets in Figure~\ref{fig:acoustophoresis-large-trajectories}. This behavior is consistent with the Bjerknes-force-based interpretation of EMB dynamics \cite{Bjerknes1906}: away from pressure antinodes, the instantaneous pressure gradient oscillates, producing an effective oscillating buoyancy that drives COM displacement. At the pressure antinode, the instantaneous gradient vanishes, eliminating buoyancy and thus halting COM motion.

\needspace{5\baselineskip}
\section{Methods}
\addcontentsline{toc}{section}{Methods}
\label{sec:methods}

An integral component of our virtual US machine is a particle‑based fluid simulation method. Such a method must satisfy several key requirements: it should remain numerically stable over short and long simulation times, be thermodynamically consistent, and accurately reproduce the fundamental hydrodynamic properties of the target fluid, in particular its viscosity and speed of sound, which are essential for the correct description of acoustic wave propagation. The method should also be coarse-grained (CG) to avoid the high costs of an all-atom simulation. CG particle-based fluid simulation methods represent multiple fluid molecules with a single CG particle, where the number of molecules per particle—known as the mapping number, $N_m$—characterizes the level of coarse-graining. Methods such as Martini, DPD, and MDPD are commonly used for mesoscale fluid simulations \cite{Lah2025}. However, as the mapping number increases, these approaches face significant challenges, including numerical freezing and instabilities. Specifically, in DPD water, crystallization occurs once the mapping number approaches 20—provided that correct physical viscosity and compressibility are enforced simultaneously \cite{Pivkin2006, Trofimov2003}. This freezing is linked to the Kirkwood–Alder transition \cite{Dzwinel2000}, observed as mapping number and speed of sound increase. The root cause is entropy loss: coarse-graining removes internal degrees of freedom, shifting free-energy minima so that the system may exhibit a premature phase transition to a frozen state even at conditions where the original atomistic system is liquid \cite{Sokhan2023,Kidder2021}.

Another well-known model for CG particle-based simulations is Martini~3 \cite{Souza2021}, which typically represents four water molecules with a single particle. Martini~3 water consists of particles that interact via pairwise Lennard-Jones interactions, tuned to reproduce the properties of water. This model is also plagued by freezing artifacts \cite{Lah2025}, which necessitate the use of so-called "antifreeze particles" \cite{Marrink2007}. At the nanoscale, alternative methods such as DPD and MDPD have been shown to better reproduce the properties of water required for US simulations \cite{Lah2025}.

The many-body extension of DPD (MDPD) allows higher CG levels while maintaining the correct compressibility. However, stability requirements force a reduction of the timestep, which negates most of the computational advantages at CG levels beyond $N_m = 10^3$ (in our simulations, we achieve $N_m = 10^6$) \cite{Trofimov2003}. Thus, achieving sufficiently high CG levels for US simulations requires a different approach.

Yet another particle-based method commonly used in fluid mechanics simulations is smoothed particle hydrodynamics (SPH) \cite{Monaghan1992}. This method differs substantially from DPD-based methods in that it is derived as a discretization of the Navier–Stokes equations (a top-down approach), unlike the DPD method, which is neither a top-down or bottom-up method \cite{Kulkarni2013}. Continuum hydrodynamics has been shown to persist down to surprisingly small scales, enabling the application of continuum theory even at the mesoscale \cite{Alder1970}. At such scales, however, molecular discreteness becomes significant, as particles exhibit thermal motion that is negligible in macroscopic systems. SPH does not account for thermal fluctuations, which restricts its applicability to macroscopic and upper-mesoscale systems. To address this limitation, SPH has been modified to include thermal fluctuations consistent with the first and second laws of thermodynamics, introducing the method known as smoothed dissipative particle dynamics (SDPD) \cite{Espanol2003}. This approach later became widely adopted for particle-based simulations at the mesoscopic scale. In its original form, however, SDPD is not suitable for simulations of US propagation at the mesoscopic scale due to the “tensile instability” artifact, which causes the fluid to fracture prematurely under tensile stress.
In the following sections, we introduce the extended version of SDPD, known as usSDPD, designed to address the tensile instability artifact in the simulations of US propagation in weakly-compressible fluids.

The simulations are performed using LAMMPS \cite{Thompson2022}. While SDPD is already implemented in LAMMPS, we extended it by adding an implicit compressible pressure solver and packaged it into the package \texttt{ICSDPD}. Additional modifications to improve tensile instability can be optionally enabled or disabled in the code.

\subsection{Smoothed dissipative particle dynamics (SDPD)}
\addcontentsline{toc}{subsection}{SDPD}
\label{sec:sdpd}

SDPD is a particle-based method for simulating hydrodynamics with thermal fluctuations, combining elements of SPH and DPD \cite{Espanol2003, Ellero2018}. The main advantage of SDPD over SPH is that it incorporates thermal fluctuations in a manner consistent with the first and second laws of thermodynamics. Compared to DPD, SDPD offers several additional benefits. It allows a general equation of state to be specified as input, rather than being limited to the fixed quadratic equation of state used in DPD. Moreover, physical parameters such as viscosity and speed of sound can be specified directly, rather than being inferred empirically from non-physical DPD parameters. Finally, mapping the system to a physical scale is straightforward, in contrast to DPD.

For an isothermal fluid, the governing equations read:
\begin{align}
    \frac{\mathrm{d} \mathbf{r}_i}{\mathrm{d}t} &= \mathbf{v}_i \nonumber \\
    \frac{\mathrm{d} \rho_i}{\mathrm{d}t} &= - \sum_j m_j \mathbf{v}_{ij} \cdot w_{ij}' \hat{\mathbf{e}}_{ij} \nonumber \\
    m_i \frac{\mathrm{d} \mathbf{v}_i}{\mathrm{d}t} &= \frac{5\eta}{3} \sum_j \frac{m_i m_j}{\rho_i \rho_j} \frac{w_{ij}'}{r_{ij}} 
    \left( \mathbf{v}_{ij} + \hat{\mathbf{e}}_{ij} \hat{\mathbf{e}}_{ij} \cdot \mathbf{v}_{ij} \right) 
    \label{eqn:SDPD-equation-of-motion} \\
    &\quad - \sum_j m_i m_j \left( \frac{p_i}{\rho_i^2} + \frac{p_j}{\rho_j^2} \right) w_{ij}' \hat{\mathbf{e}}_{ij} \notag \\
    &\quad + \sum_j \sqrt{\frac{20\eta}{3} k_B T \frac{m_i m_j}{\rho_i \rho_j} w_{ij}' }
    \frac{\mathrm{d} \mathsf{\overline W}_i}{\mathrm{d}t} \cdot \hat{\mathbf{e}}_{ij}, \notag
\end{align}
where $\mathbf{r}_i$, $\mathbf{v}_i$, $m_i$, $\rho_i$, and $p_i$ denote the position, velocity, mass, density, and pressure of the $i$-th particle, respectively. $\hat{\mathbf{e}}_{ij} = \mathbf{r}_{ij}/r_{ij}$ is the unit vector along $\mathbf{r}_{ij} \equiv \mathbf{r}_j - \mathbf{r}_i$, $\mathbf{v}_{ij} \equiv \mathbf{v}_j - \mathbf{v}_i$, and $w_{ij}'$ is the magnitude of the kernel function gradient, such that $\nabla_i W_{ij} = w_{ij}' \hat{\mathbf{e}}_{ij}$, with $W_{ij} \equiv W(\mathbf{r}_j - \mathbf{r}_i)$. In all simulations, we use the cubic spline kernel and neglect the bulk viscosity.
$\mathrm{d} \mathsf{\overline W}_i$ is the symmetric part of a $3 \times 3$ matrix of independent Wiener process increments. The independent Wiener process increments can be simulated as $\zeta (\Delta t)^{1/2}$, where $\zeta$ is an independent Gaussian random number and $\Delta t$ is the timestep. Further details are provided in \cite{Jalalvand2020, Espanol2003}.

In SPH, it is common to use an alternative form of the continuity equation:
\begin{equation}
    \rho_i = \sum_j m_j W_{ij}.
    \label{eqn:density-summation}
\end{equation}
In our SDPD modifications, described in Sections \ref{sec:icSDPD} and \ref{sec:improving-icsdpd-at-negative-pressures}, we also adopt Equation (\ref{eqn:density-summation}) to compute densities, as this formulation simplifies the implementation of boundary conditions (see the code for details).
We additionally employ the standard (leapfrog) SPH integrator, which is already available in the SPH package of LAMMPS. The procedure is summarized in Algorithm \ref{alg:sph-integrator}.

\begin{algorithm}
\caption{Smoothed particle hydrodynamics time integration scheme}
\label{alg:sph-integrator}
\begin{algorithmic}[1]

\Procedure{Pre-force computations}{}
    \ForAll{particles $i$}
        \State compute half-step velocity: $\mathbf{v}_i^{t+\Delta t/2} = \mathbf{v}_i^t + \tfrac{\Delta t}{2 m_i} \mathbf{f}_i^t$
        \State compute predicted velocity: $\tilde{\mathbf{v}}_i^{t+\Delta t} = \mathbf{v}_i^t + \tfrac{\Delta t}{m_i} \mathbf{f}_i^t$
        \State compute position: $\mathbf{r}_i^{t+\Delta t} = \mathbf{r}_i^t + \Delta t \,\mathbf{v}_i^{t+\Delta t/2}$
    \EndFor
\EndProcedure

\Procedure{Force computation}{}
    \ForAll{particles $i$}
        \State compute new forces $\mathbf{f}_i^{t+\Delta t}$
    \EndFor
\EndProcedure

\Procedure{Post-force computations}{}
    \ForAll{particles $i$}
        \State compute full-step velocity: $\mathbf{v}_i^{t+\Delta t} = \mathbf{v}_i^{t+\Delta t/2} + \tfrac{\Delta t}{2 m_i}\mathbf{f}_i^{t+\Delta t}$
    \EndFor
\EndProcedure

\end{algorithmic}
\end{algorithm}

In SDPD, scaling length and time by 10 reduces dimensionless temperature by 1000 and dimensionless viscosity by 10, with speed of sound unchanged (see Section \ref{sec:scaling-of-dimensionless-quantities}). An important consequence is that, at larger scales, the gap between the dimensionless speed of sound and the dimensionless viscosity widens, ultimately limiting the ability to simulate water at very high coarse-graining levels. This limitation is illustrated in the Supplementary Information, Section \ref{sec:analysis-of-the-fluid-in-equilibrium} and has also been mentioned in \cite{Lah2025}. A numerical value that quantifies the discrepancy between viscous and sonic time scales is the so-called compressibility factor (the ratio of the time for sound to travel one particle radius $a$ to the time for viscous diffusion over the same distance) \cite{Tatsumi2013}
\begin{equation}
    \varepsilon \equiv \frac{a/c}{a^2/\nu} = \frac{\nu}{a c}.
\end{equation}
Moreover, the reduction of dimensionless temperature by a factor of $1000$ with each 10-fold increase in scale implies that thermal fluctuations rapidly become negligible compared to other forces when transitioning from the nanometer to the micrometer scale. For most simulations above the micrometer scale, thermal fluctuations should therefore be disabled. In the absence of thermal fluctuations, SDPD reduces to the well-known smoothed particle hydrodynamics (SPH) method, commonly employed for macroscopic particle-based hydrodynamic simulations.

\subsection{Implicit-compressible pressure solver}
\label{sec:icSDPD}

In weakly compressible fluids, the speed of sound greatly exceeds typical advective velocities, causing the timestep to be severely constrained by acoustic propagation. To alleviate this restriction, we evaluate pressure forces using an implicit scheme, while dissipative and random forces are treated explicitly \cite{Hu2004,Ikari2005,Khayyer2009,Solenthaler2009,Gissler2020}. This hybrid approach permits timesteps up to 40 times larger than those of the original SDPD method (see Supplementary Information, Section~\ref{sec:analysis-of-the-fluid-in-equilibrium}).

\begin{figure}[h!]
    \centering
    \includegraphics[width=0.75\textwidth]{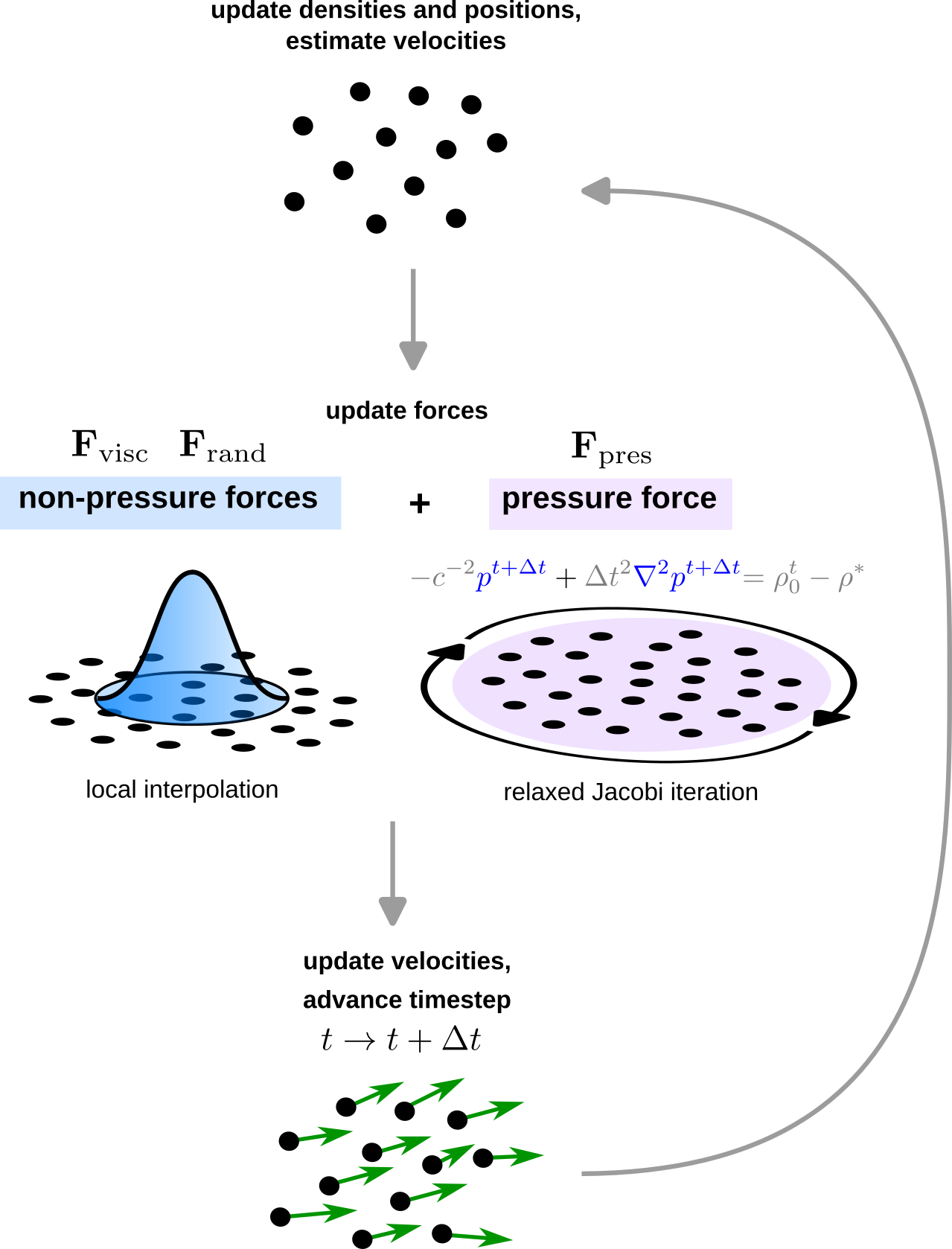}
    \caption{\textbf{Graphical representation of the timestepping procedure of the SDPD method with the implicit compressible pressure solver.} For details, see Algorithm~\ref{alg:sph-integrator} and Algorithm~\ref{alg:pressure-solver}. First, the densities and positions of the particles are updated, and the velocities at the next timestep are estimated. These quantities are then used to update the forces. The so-called non-pressure forces (viscous and random contributions) are computed locally from particle properties. Subsequently, a global iteration procedure is carried out to compute the pressure for every particle, which is then used to determine the pressure force. Finally, the sum of the non-pressure and pressure forces is used to update the particle velocities and advance the timestep.}
\label{fig:icsdpd-method}
\end{figure}

An implicit pressure solver for compressible flows was originally developed within the SPH framework for snow simulations \cite{Gissler2020} and later incorporated into the open-source SPH solver SPlisHSPlasH \cite{Bender2025}. Since SDPD shares the same kernel-based discretization principles as SPH, this pressure solver can be naturally adapted to the SDPD framework.
To derive the implicit compressible pressure solver, we begin from the continuity and Navier–Stokes equations,
\begin{align*}
    &\frac{\mathrm{D}\rho}{\mathrm{D}t} + \rho \nabla \cdot \mathbf{v} = 0 \\
    \rho &\frac{\mathrm{D}\mathbf{v}}{\mathrm{D}t} = -\nabla p + \rho \mathbf{a}^\textrm{non-pressure},
\end{align*}
where $\mathbf{a}^\textrm{non-pressure}$ includes all non-pressure forces acting on the fluid. After discretizing the equations, doing some simplifications and assuming a linear EOS, we get \cite{Gissler2020,Hu2004,Ikari2005,Khayyer2009}
\begin{equation}
	-c^{-2}p^{t+\Delta t} + \Delta t^2 \nabla^2 p^{t+\Delta t} = \rho_0^t - \rho^*,
	\label{eqn:implicit-eqn}
\end{equation}
where $c$ is the speed of sound, $\rho^{*} = \rho^t - \Delta t \rho^t \nabla \cdot \mathbf{v}^*$ and $\mathbf{v}^* = \mathbf{v}^t + \Delta t \mathbf{a}^\textrm{non-pressure}$. Here, $\rho^t$ and $\mathbf{v}^t$ are the density and velocity at timestep $t$, and $\mathbf{a}^\textrm{non-pressure}$ denotes the contribution to the fluid particle acceleration from dissipative and random forces. $V_i = m_i/\rho_i$ is the volume of the $i$-th particle. The summations in the algorithm are performed over fluid particles (indexed by $j$), boundary particles (indexed by $b$), or both (indexed by $k$). The implicit compressible pressure solver employs the relaxed Jacobi iteration to solve Equation~(\ref{eqn:implicit-eqn}) iteratively; iteration step is denoted with the superscript $(\ell)$. Algorithm~\ref{alg:pressure-solver} provides a slightly modified version of the procedure described in \cite{Gissler2020}. In the algorithm, $\psi = 1.5$, and $\omega = 0.5$. A graphical overview of the timestepping procedure is shown in Figure~\ref{fig:icsdpd-method}.
The implicit compressible pressure solver iterates until the density error falls below \SI{0.1}{\percent}, i.e. when $\frac{1}{N} \sum_i \left ( \rho_0 - \rho_i^* - (A p)^{(\ell)}_i \right ) < 0.001 \rho_0$, where $N$ is the total number of fluid particles.

\begin{algorithm}
\caption{Solver steps of the implicit compressible pressure solver.}
\label{alg:pressure-solver}
\begin{algorithmic}[1]
\Procedure{Prepare}{}
    \ForAll{particles $i$}
    	\State compute $\mathbf{v}_i^* = \mathbf{v}_i^t + \Delta t \mathbf{a}_i^\textrm{non-pressure}$
    	\State compute $\nabla \cdot \mathbf{v}_i^* = \sum_k (\mathbf{v}_k^* - \mathbf{v}_i^*) V_k \nabla W_{ik}$
        \State compute $\rho_i^{*} = \rho_i^t - \Delta t \rho_i^t \nabla \cdot \mathbf{v}_i^*$
        \State compute $$a_{ii} = -c^{-2} - \Delta t^2 \sum_j V_i V_j \|\nabla W_{ij}\|^2
- \Delta t^2 \left( \sum_j V_j \nabla W_{ij} + \psi \sum_b V_b \nabla W_{ib} \right) \sum_k V_k \nabla W_{ik}$$
    \EndFor
\EndProcedure

\Procedure{Solve}{}
    \While{not converged}
        \ForAll{particles $i$}
            \State compute $\nabla p_i^{(\ell)} = \sum_j (p_j^{(\ell)} + p_i^{(\ell)}) V_j \nabla W_{ij} + \psi p_i^{(\ell)} \sum_b V_b \nabla W_{ib}$
        \EndFor
        \ForAll{particles $i$}
        	\State compute $\nabla^2 p_i^{(\ell)} = \sum_j (\nabla p_j^{(\ell)} + \nabla p_i^{(\ell)}) V_j \nabla W_{ij}$
            \State compute $(A p)^{(\ell)}_i = -c^{-2} p^{(\ell)}_i + \Delta t^2 \nabla^2 p^{(\ell)}_i$
            \State compute $p_i^{(\ell+1)} = p_i^{(\ell)} + \frac{\omega}{a_{ii}} \left( \rho_0 - \rho_i^* - (A p)^{(\ell)}_i \right)$
        \EndFor
    \EndWhile
\EndProcedure

In the SOLVE procedure, index $t+\Delta t$ is removed for clarity.

\end{algorithmic}
\end{algorithm}

\subsection{Providing stability at negative pressures (usSDPD)}
\label{sec:improving-icsdpd-at-negative-pressures}

In SPH, SDPD, and related particle-based methods, pressure forces originate from the pressure term in the equations of motion. Around zero pressure, these forces are minimal, but their magnitude increases with both positive and negative pressures. In an equilibrium fluid, large pressure forces acting on individual particles largely cancel, resulting in a relatively small net force. However, any inherent instabilities in the method become strongly pronounced at high positive or negative pressures. The extent of compression or expansion at which instabilities inhibit simulations depends on the rate at which pressure changes with density, i.e., the speed of sound. This explains why the same methods perform better in fluids with higher compressibility (lower speed of sound) than in those with lower compressibility (higher speed of sound).
One such instability is tensile instability, which manifests as premature fluid fracture under negative pressures. This instability poses a significant challenge for simulating US propagation, as acoustic waves involve alternating regions of high and low pressure; in the low-pressure regions, tensile instability can emerge and disrupt the simulation. We observe that many existing remedies for tensile instability make the fluid slightly viscoplastic (introduce some spurious elasticity) at highly positive/negative pressures \cite{Takeda1994,Randles2000,Sugiura2016,Sigalotti2006}. While such modifications may be acceptable for certain systems, they are unsuitable for studies of acoustic radiation forces, as in the present work. These artifacts become particularly pronounced at larger scales—where thermal fluctuations are negligible—and in low Mach number flows, where convective time scales are much slower than acoustic time scales. In some cases, these spurious elastic forces can even lead to freezing phenomena at high positive or negative pressures.

\begin{figure}[h!]
    \centering
    \includegraphics[width=0.5\textwidth]{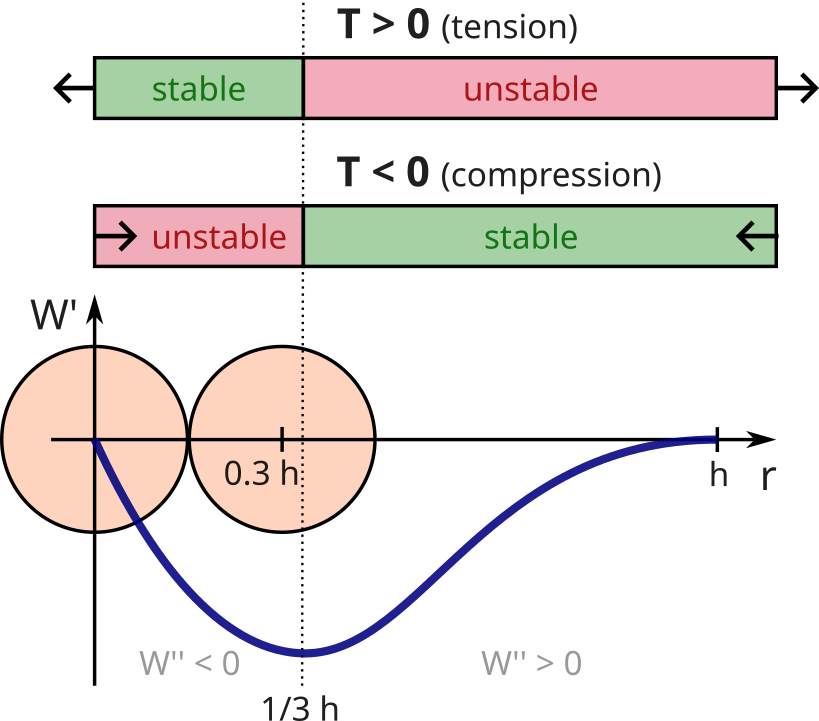}
    \caption{\textbf{Stable and unstable regimes of the SPH-based methods.} First derivative of a cubic spline kernel (blue) along with regions of pair and tensile instabilities (red). The derivative of the cubic spline kernel has a minimum at $r = 1/3 h$. Two particles are also shown to scale (orange) to illustrate that at zero background pressure, immediate neighbors are in the pair instability regime if $a=0.3 \, h$. For the conventional choice of $a=0.5 \, h$, immediate neighbors are in the tensile instability regime.}
    \label{fig:tensile-instability-kernel}
\end{figure}

It can be shown, following Swegle \cite{Swegle1995}, that tensile instability arises when $W''T > 0$, where $W''$ is the second derivative of the kernel with respect to its argument, and $T$ is the tensile stress (positive for tension and negative for compression). From this condition, it follows that a region stable under tension is unstable under compression, and vice versa. The condition is illustrated in Figure \ref{fig:tensile-instability-kernel}.

To mitigate instabilities at high negative pressures, we introduce two additional modifications. First, we reduce the fluid particle diameter from $d_p = 0.5 \, h$ to $0.3 \, h$. As a result, the typical number of neighbors increases from approximately $60$ to about $250$, slightly raising the computational cost. More importantly, some neighbor particles are then forced into the pair-instability regime, where particles form fully overlapping pairs with their neighbors \cite{Swegle1995}. To avoid this issue, an additional Lennard-Jones (LJ) potential is introduced with a small, empirically-determined well depth of $\varepsilon = \SI{1e-21}{J}$ and $\sigma = 0.23 \, h$, which is smaller than the particle diameter of $0.3 \, h$. This potential has a minimum at $r_{ij} \approx 0.26 \, h$. The purpose of this repulsion is to prevent pair instability when particles are unphysically close, while leaving the system unaffected when particles are reasonably separated. This means that at sustained positive pressures, fluid particles tend to cluster at some slightly smaller interparticle distance, determined by the LJ potential, however, the effect is weak and reversible, when pressures are lower. The idea of adding a LJ potential to the SPH method was first proposed by Monaghan \cite{Monaghan2000}, but was not implemented in that work. Although this approach is effective in our case, many potential improvements to the simple repulsive LJ potential remain possible. As demonstrated in the Supplementary Information, Section \ref{sec:analysis-of-the-fluid-in-equilibrium}, the LJ potential introduces an additional viscosity contribution, which constrains the achievable coarse-graining level in our simulations. Improving the repulsive potential is left for future work.

The second modification incorporates Monaghan’s “artificial pressure” to further improve stability at negative pressures \cite{Monaghan2000}.
The main idea of artificial pressure is to reduce the magnitude of interparticle pressure forces under negative pressure, thereby shifting the onset of tensile instability toward lower pressures.
To implement artificial pressure, we add an additional term $R f_{ij}^n$ to Equation~(\ref{eqn:SDPD-equation-of-motion}), as follows
\begin{equation}
	\left( \frac{p_i}{\rho_i^2} + \frac{p_j}{\rho_j^2} \right) \to 
	\left( \frac{p_i}{\rho_i^2} + \frac{p_j}{\rho_j^2} + R f_{ij}^n \right),
	\label{eqn:artificial-pressure}
\end{equation}
where $f_{ij} = W(r_{ij})/W(r_{\mathrm{ave}})$ and $r_{\mathrm{ave}}$ is the average spacing between fluid particles (set to $0.3 \, h$ in all simulations with artificial pressure). The factor $R$ is computed using the procedure in (\ref{alg:artificial-pressure}), and $n$ is set to $4$.
\begin{algorithm}
    \caption{Computation of Artificial Pressure}
	\label{alg:artificial-pressure}
    \begin{algorithmic}[1]
        \Procedure{ComputeArtificialPressure}{}
        	\State $R_i = \max \left ( -\frac{\epsilon p_i}{\rho_i^2}, 0 \right )$
        	\State $R_j = \max \left ( -\frac{\epsilon p_j}{\rho_j^2}, 0 \right )$
            \If{$p_i > 0$ \textbf{and} $p_j > 0$}
                \State $R = 0.01 \left( \frac{p_i}{\rho_i^2} + \frac{p_j}{\rho_j^2} \right)$
            \Else
                \State $R = R_i + R_j$
            \EndIf
        \EndProcedure
    \end{algorithmic}
    $\epsilon$ is set to $0.3$ in all simulations that use artificial pressure.
\end{algorithm}
We note that under negative pressures, expression (\ref{eqn:artificial-pressure}) simplifies to
\begin{align}
	\frac{p_i}{\rho_i^2} + \frac{p_j}{\rho_j^2} + R f_{ij}^n &= 
	\frac{p_i}{\rho_i^2} + \frac{p_j}{\rho_j^2} - \epsilon f_{ij}^n \left ( \frac{p_i}{\rho_i^2} + \frac{p_j}{\rho_j^2} \right )	\\
	&= (1-\epsilon f_{ij}^n) \left ( \frac{p_i}{\rho_i^2} + \frac{p_j}{\rho_j^2} \right ),
\end{align}
which clearly illustrates the artificial reduction of pressure forces between fluid particles in the negative-pressure regime.

\begin{figure}[h!]
    \centering
    \includegraphics[width=0.55\linewidth]{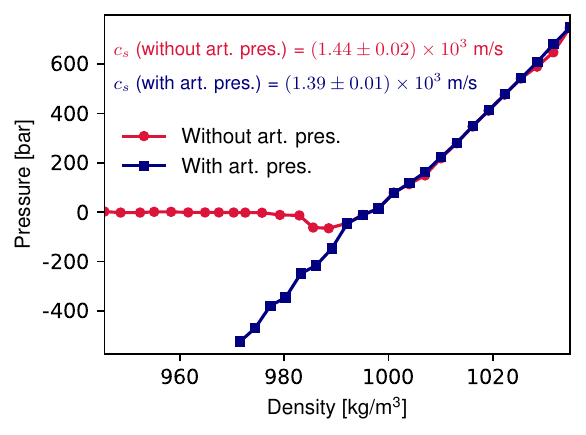}
    \caption{\textbf{Pressure-density relationship (EOS) for usSDPD with and without artificial pressure.} Both simulations use a linear EOS. Without artificial pressure, the simulated EOS matches the input only at positive pressures; at negative pressures, a divergence caused by tensile instability becomes evident. In contrast, the method with artificial pressure accurately reproduces a linear EOS even under highly negative pressures. Simulations are performed on a \SI{0.1}{\um} granularity with viscosity $\eta = \SI{8.9e-4}{\pascal \second}$ and speed of sound $c_s = \SI{1481}{m/s}$ at $T = \SI{300}{K}$. Both pressure and density are measured using the non-biased approach introduced in \cite{Kulkarni2013}. The inset shows the speed of sound, obtained from the slope of the linear region of the EOS curve.}
    \label{fig:eos}
\end{figure}

We further verify that usSDPD remains stable across a wide pressure range by examining the equation of state (EOS), which is linear in our case; $p_i = \rho_0 c^2 \left ( \rho_i / \rho_0 - 1 \right )$, where $\rho_0$ is the prescribed input density, $c$ is the speed of sound, and $\rho_i$ and $p_i$ are the density and pressure of the $i$-th fluid particle. By definition, the pressure is zero when $\rho = \rho_0$. The speed of sound is obtained from the slope of the $p(\rho)$ curve using the standard relation $c^2 = \partial p / \partial \rho$, making EOS plots a direct means of verifying the speed of sound (compressibility) in the simulation. As shown in Figure \ref{fig:eos}, the use of usSDPD substantially extends the range of stable negative pressures. At densities higher than the reference density (positive pressures), the EOS is linear as expected in both cases. However, at lower densities, the method without artificial pressure diverges from the linear relation, while the method with artificial pressure stays linear even for highly negative pressures.
At very high tensions, we note that a small amount of spurious elasticity remains, but its magnitude is sufficiently low to be acceptable for our purposes.

\subsection{Open boundary molecular dynamics (OBMD)}
\label{sec:OBMD}

OBMD is a simulation framework designed to model nonequilibrium systems by enabling the controlled exchange of mass, momentum, and energy across the boundaries of a finite simulation domain \cite{Buscalioni2015}. We use OBMD to model the transducers in the virtual US machine. The framework partitions the system into a central region of interest (ROI) and adjacent buffer zones, where particles are inserted or removed. The particle number in each buffer is regulated by a feedback mechanism that ensures convergence toward a target density, while particle insertion is performed using energy-minimizing algorithms such as USHER \cite{Buscalioni2003}. Boundary conditions are imposed by applying external forces to buffer particles, derived from a local momentum balance that accounts for both interparticle interactions and the momentum flux associated with particle exchange. This approach allows for the imposition of stress or velocity boundary conditions.

In order to correctly model the constant chemical potential of the system, the change in energy must be calculated when a particle is inserted at some position in the buffer. In simpler pairwise fluid methods, the energy increment is obtained by summing the additional pairwise contributions to the total energy for every neighbor of the inserted particle. In multibody, non-local methods such as the one used in this work, the exact calculation of the energy increment is more difficult, as the non-local effect of adding a particle on the pressure must be considered. Therefore, we employ a crude approximation, which is sufficient for the simple cases presented in this paper.
To obtain the approximate energy increment, the density of the inserted particle is computed from the positions of its neighbors. The pressure of the inserted particle is then determined from its density using a simple Cole equation of state, instead of solving the pressure iteratively:
\begin{equation}
	p_i = B \left [ \left ( \frac{\rho_i}{\rho_0} \right )^\gamma -1 \right ],
\end{equation}
where $B=\rho_0 c^2/\gamma$, $\rho_0$ is the equilibrium density, $c$ is the speed of sound, and $\gamma = 7$. The Cole equation is a standard choice for simulating water with SDPD, however, we could have as well chosen the linear EOS, since we use it in the implicit compressible pressure solver and the Cole EOS is approximately linear at the range of pressures we are working with. The approximate energy increment is then calculated as
\begin{equation}
	\Delta E \approx p_i V_i - \frac{1}{2} \sum_j m_i m_j \left ( \frac{p_i}{\rho_i^2} + \frac{p_j}{\rho_j^2} \right ) \mathbf{v}_{ij} \cdot \nabla_j W_{ij},
\end{equation} 
where $V_i = m_i/\rho_i$. The first term on the right-hand side corresponds to the pressure work associated with creating a particle of volume $V_i$ at pressure $p_i$, while the second term accounts for the energy increments from the SPH energy equation (Equation 3.17 in \cite{Monaghan1992}).

\needspace{5\baselineskip}
\section{Conclusion}

In this work, we developed a virtual US machine capable of simulating US-fluid-structure interaction between a weakly-compressible fluid and immersed structures at mesoscopic (micrometer) scales, corresponding to medically-relevant MHz–GHz frequency range. We achieved this by modifying the SDPD method to accurately reproduce acoustic wave propagation with realistic speed of sound and viscosity. By introducing an implicit compressible pressure solver, we extended the timestep by 40  times compared to explicit schemes—a significant improvement for weakly compressible fluids such as water.
We also improved the EOS curves (eliminated tensile instability) by reducing the fluid particle diameter, introducing artificial pressure and adding additional Lennard-Jones interaction. This improved the stability of the US simulation at negative pressures, which is an essential part of any virtual US machine.
Because the fluid simulation framework presented here is particle-based, it can be directly coupled with particle-based descriptions of immersed structures. As a proof of concept, we coupled the method with an elastic EMB filled with DPD particles and subjected it to US excitation. The coupled simulation successfully reproduced the expected migration of the EMB to the pressure antinode.

Some interesting research questions have been left for future investigation. 
For simplicity, the bulk viscosity of the surrounding fluid was set to zero, even though it strongly influences US attenuation. Consequently, the attenuation predicted by our simulations is not expected to match experimental observations. Since bulk viscosity is already included in the general SDPD framework, extending usSDPD to account for it will be straightforward. 
Furthermore, the role of alternative interparticle repulsive interactions beyond the Lennard-Jones potential remains unexplored; a more carefully designed repulsion may reduce viscosity artifacts while still preventing pair instability. 
Another direction for future work is the detailed investigation of fluid–membrane interactions, which were not addressed in this study.
Finally, to extend the applicability of our virtual US machine, the implementation of non-reflective boundary conditions (NRBCs) will be essential. While we restricted ourselves to standing wave simulations, NRBCs are necessary to accurately model traveling waves and complex pulse sequences. Several promising approaches for NRBCs have been proposed in related hydrodynamics and SPH contexts, including methods based on the method of characteristics \cite{Lastiwka2009}, fluid-property extrapolation \cite{Tafuni2018}, and particle-property freezing \cite{Khorasanizade2016}. Adapting these techniques to our usSDPD-based framework will enable simulations with arbitrary US pulses, moving beyond standing waves. This capability is particularly important for realistic experimental scenarios, where systems are typically insonicated with short, high-intensity pulses to mitigate sample heating \cite{Cheng2019}. 
Although heating effects were not included in our isothermal formulation, the extension to a thermal US machine is feasible, since thermal effects are already incorporated in the most general SDPD method \cite{Espanol2003}.
The usSDPD method also presents certain limitations. At positive pressures, the method gradually develops pair instability, leading to particle clustering at the interaction range of the Lennard-Jones potential. While negligible under oscillatory US conditions, this effect becomes relevant under sustained positive pressures. We speculate that adaptive kernels \cite{Monaghan1992} could resolve this issue, which we leave for future work. At highly negative pressures, the method also exhibit some spurious elasticity.

In conclusion, the advances presented in this work establish a foundation for particle-based simulations of US propagation in mesoscopic systems. By combining improved numerical stability and natural coupling with particle-based structures, the virtual US machine provides a versatile framework for investigating US–fluid–structure interactions, with promising applications ranging from studies of wave propagation to studies of various mesoscale systems, such as gas vesicles, EMBs, red blood cells, microrobots, cells, and other biologically or technologically relevant structures.

\backmatter

\bmhead{Code Availability}
The code will be made publicly available in our GitHub repository upon acceptance of the paper.

\bmhead{Acknowledgments}

The authors acknowledge financial support under the ERC Advanced Grant MULTraSonicA (Grant No.\,885155) from the European Research Council. This project has also received funding from the European High Performance Computing Joint Undertaking (Grant No.\,101093169). The authors also acknowledge the HPC RIVR consortium (www.hpc-rivr.si) and EuroHPC JU (eurohpc-ju.europa.eu) for providing computing resources of the HPC system Vega at the Institute of Information Science (www.izum.si). The authors further thank dr. Franci Merzel for insightful discussions.

\bmhead{Author contributions}

U.Č. performed the simulations and wrote the paper. T.P. and M.P. reviewed the paper. U.Č. and T.P. wrote the software. M.P. acquired the funding and designed the study.

\bmhead{Competing interests}
The authors declare no competing interests.
\newpage

\bibliography{particle-based.bib}

\clearpage
\setcounter{section}{0}
\renewcommand{\thesection}{S\arabic{section}}

\section{Supplementary Information}
\label{sec:supplementary}

\subsection{Scaling of dimensionless quantities}
\label{sec:scaling-of-dimensionless-quantities}

To ensure numerical stability and ease of use, all simulations in this study are expressed in dimensionless form. The user defines three fundamental reference scales—length, time, and density—from which all other physical quantities, such as mass, energy, and viscosity, are derived. The dimensionless temperature is expressed in terms of thermal energy as $T^* = k_B T / E_0$, where $k_B$ is the Boltzmann constant, $E_0$ is the energy scale, and the superscript asterisk denotes a dimensionless quantity.
A change in simulation scale (granularity) is implemented by multiplying both the characteristic length and time by a constant factor. Since the timestep is constrained by the time required for an acoustic signal to traverse one cutoff distance, the cutoff and the maximum stable timestep increase proportionally under such scaling. This procedure maintains a constant non-dimensional speed of sound. Typical scales, used in the article are given in Table \ref{tab:scales}.
Most terms in the non-dimensional equation of motion (\ref{eqn:SDPD-equation-of-motion}) are invariant under this scaling transformation. The only quantities that vary are the dimensionless viscosity and temperature, which adjust according to the following relations when the simulation scale is increased by a factor of ten:
\begin{align*}
	\eta_\mathrm{before}^* = \frac{\eta}{\eta_0} &\to \frac{\eta}{10 \eta_0} = \eta_\mathrm{after}^* = \frac{1}{10}\eta_\mathrm{before}^* \\
	T_\mathrm{before}^* = \frac{k_B T}{E_0} &\to \frac{k_B T}{10^3 E_0} = T_\mathrm{after}^* = \frac{1}{1000} T_\mathrm{before}^*.
\end{align*}
In these expressions, the subscripts “before” and “after” indicate values prior to and following the scaling transformation, respectively, while the subscript “0” denotes the base scale derived from the three fundamental reference scales.

\begin{table}[h]
    \centering
    \caption{\textbf{Simulation Scales (Granularities)}}
    \begin{tabular}{lc}
        \toprule
        \textbf{Length Scale} & \textbf{Time Scale} \\
        \midrule
        \SI{1e-8}{\meter} & \SI{1e-9}{\second} \\
        \SI{1e-7}{\meter} & \SI{1e-8}{\second} \\
        \SI{1e-6}{\meter} & \SI{1e-7}{\second} \\
        \bottomrule
    \end{tabular}
    \label{tab:scales}
\end{table}

\subsection{Analysis of the fluid in equilibrium}
\label{sec:analysis-of-the-fluid-in-equilibrium}

In this section, we compare the novel usSDPD method with the standard (vanilla) SDPD implementation.
To evaluate the stability of the investigated methods, we simulate a fluid in equilibrium and monitor the density of a representative particle over time, as shown in Figure~\ref{fig:singledensity}. Simulation parameters are listed in detail in the Supplementary Information, Section \ref{sec:parameters}. Figure~\ref{fig:singledensity-sdpd-2} reveals that, at a time scale of \SI{2e-9}{s}, vanilla SDPD method exhibits non-physical pressure fluctuations. These oscillations disappear only when the time scale is reduced to \SI{1e-9}{s}, as shown in Figure~\ref{fig:singledensity-sdpd}. In contrast, usSDPD shows an approximately 40 times larger stable timestep length (Figures~\ref{fig:singledensity-icsdpd-lj-artpres} and \ref{fig:singledensity-icsdpd-lj-artpres-2}).

\begin{figure}[h!]
    \centering
    \begin{subfigure}[t]{0.49\textwidth}
        \centering
        \caption{}
        \includegraphics[width=\textwidth]{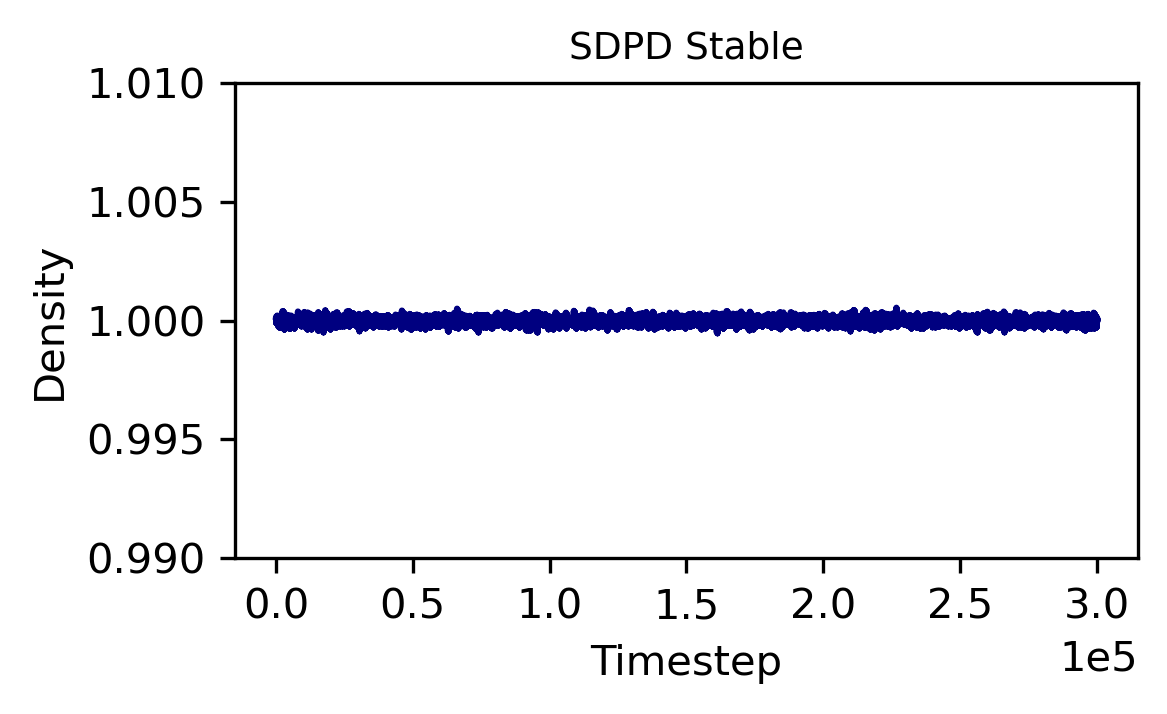}
        \label{fig:singledensity-sdpd}
    \end{subfigure}
    \hfill
    \begin{subfigure}[t]{0.49\textwidth}
        \centering
        \caption{}
        \includegraphics[width=\textwidth]{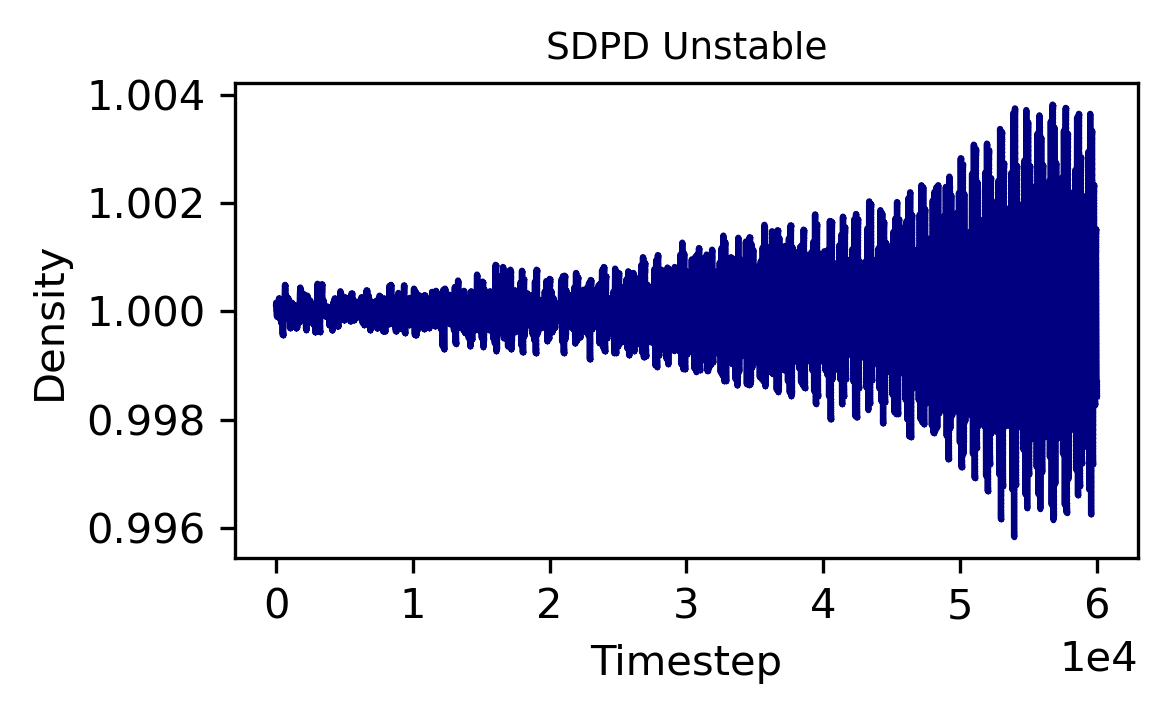}
        \label{fig:singledensity-sdpd-2}
    \end{subfigure}
    
    \vspace{-2em}
    
    \centering
    \begin{subfigure}[t]{0.49\textwidth}
        \centering
        \caption{}
        \includegraphics[width=\textwidth]{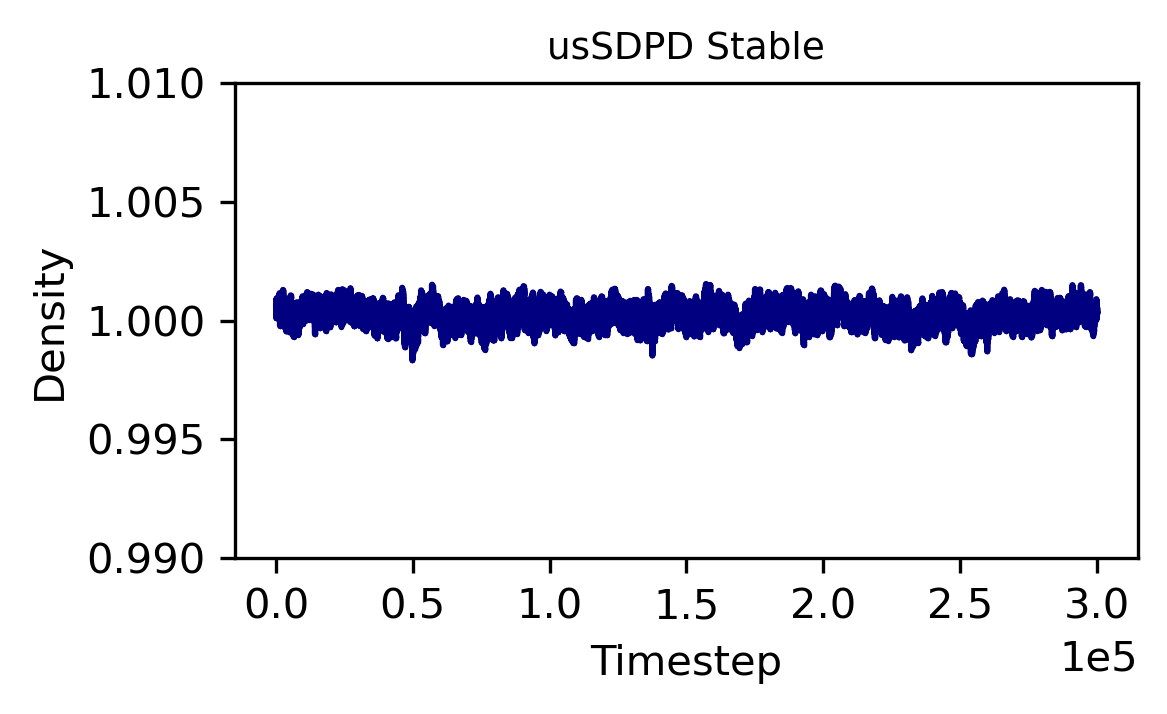}
        \label{fig:singledensity-icsdpd-lj-artpres}
    \end{subfigure}
    \hfill
    \begin{subfigure}[t]{0.49\textwidth}
        \centering
        \caption{}
        \includegraphics[width=\textwidth]{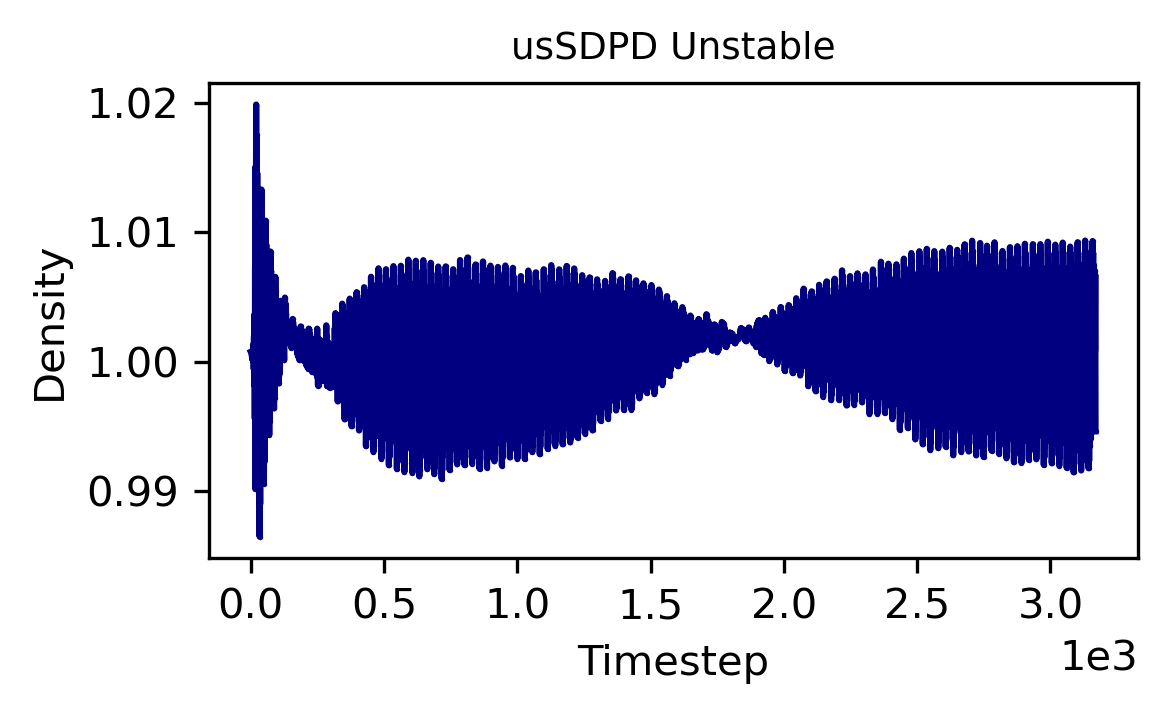}
        \label{fig:singledensity-icsdpd-lj-artpres-2}
    \end{subfigure}
    
    \caption{\textbf{Stability analysis for SDPD and usSDPD methods on a \SI{0.1}{\um} granularity.} (a) SDPD, time-scale \SI{1e-9}{s}, (b) SDPD, time-scale \SI{2e-9}{s}, (c) usSDPD, time-scale \SI{4e-8}{s}, (d) usSDPD, time-scale \SI{5e-8}{s}. The data is left in dimensionless units for clarity.}
    \label{fig:singledensity}
\end{figure}

\begin{figure}[h!]
    \centering
    \begin{subfigure}[t]{0.49\textwidth}
        \centering
        \caption{}
        \includegraphics[width=\textwidth]{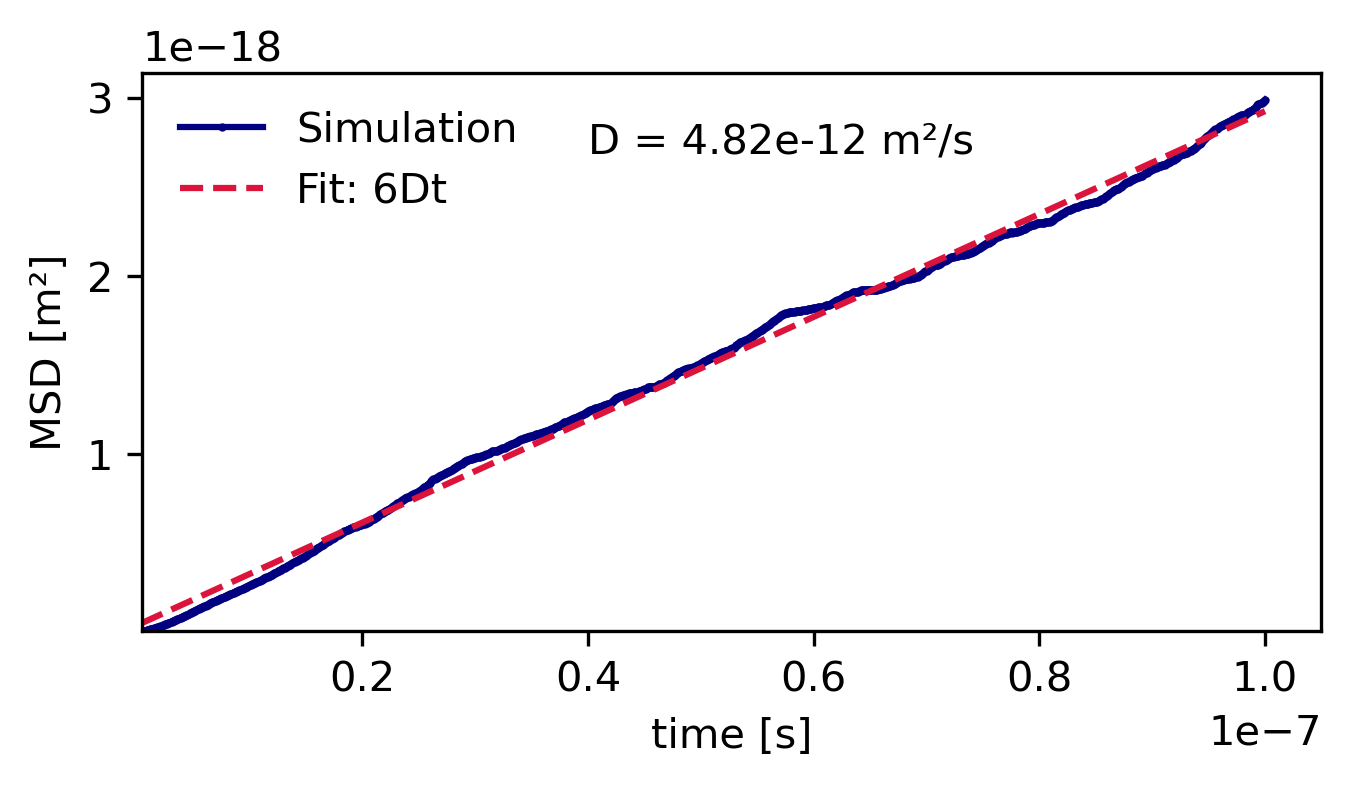}
        \label{fig:msd-sdpd}
    \end{subfigure}
    \hfill
    \begin{subfigure}[t]{0.49\textwidth}
        \centering
        \caption{}
        \includegraphics[width=\textwidth]{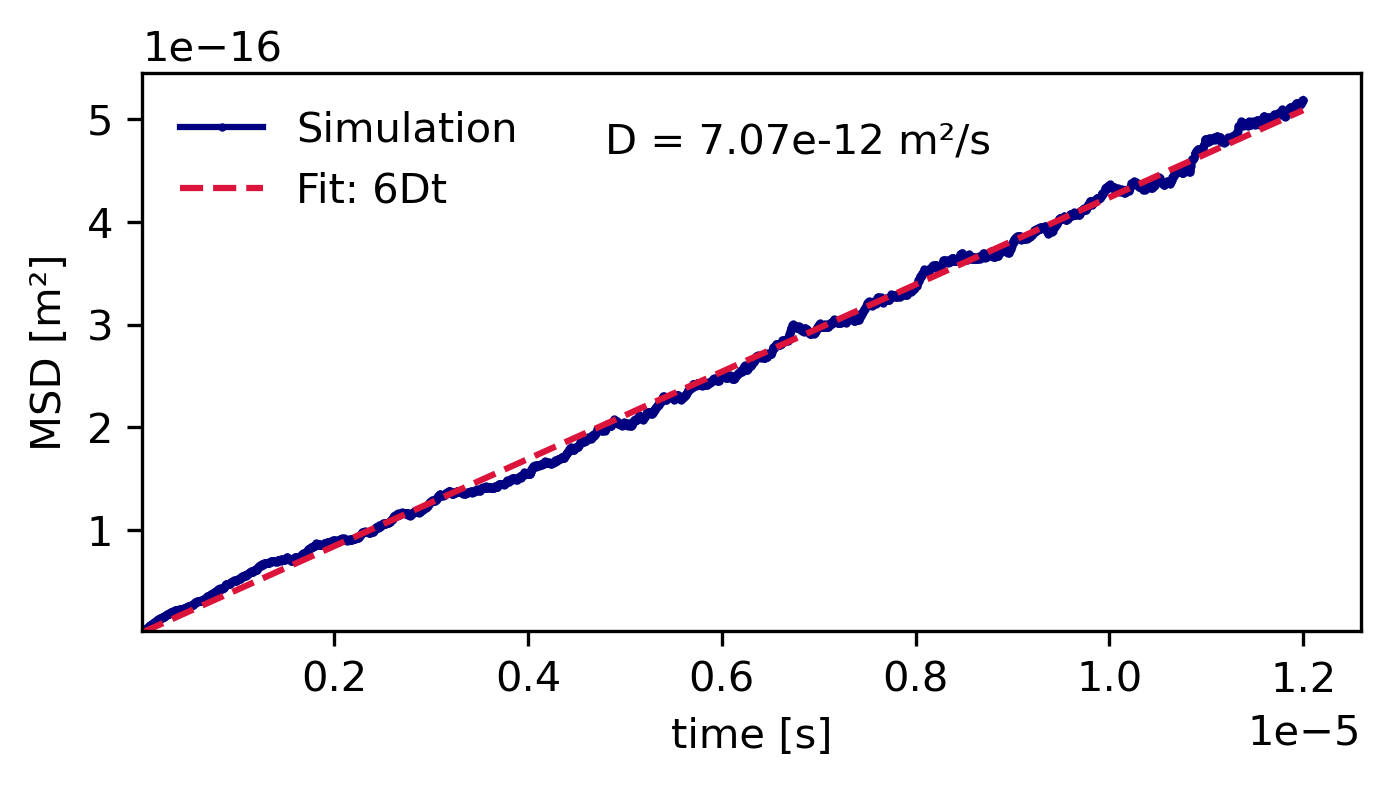}
        \label{fig:msd-icsdpd-lj-artpres}
    \end{subfigure}
    
    \vspace{-1em}

    \begin{subfigure}[t]{0.49\textwidth}
        \centering
        \caption{}
        \includegraphics[width=\textwidth]{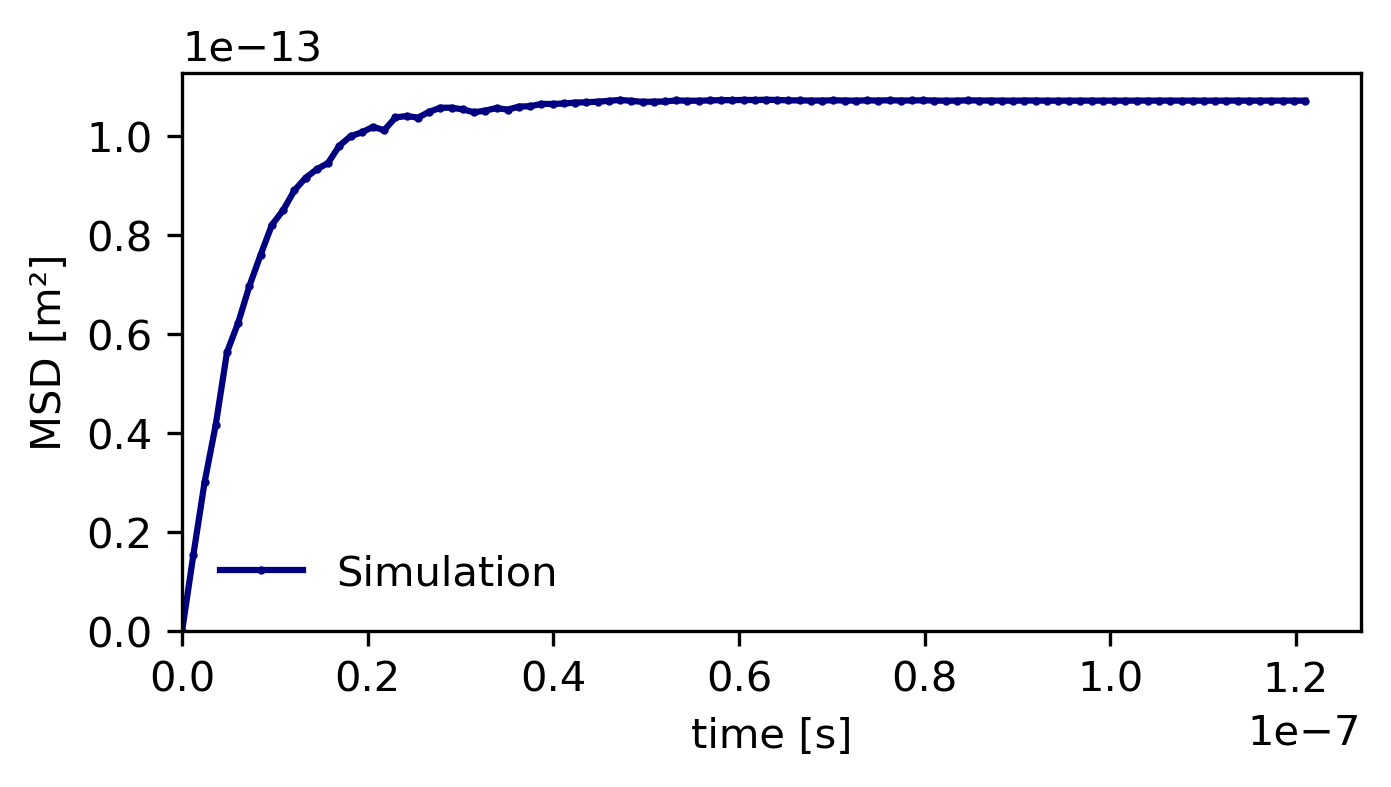}
        \label{fig:msd-dpd}
    \end{subfigure}
    
    \caption{\textbf{MSD curves for (a) vanilla SDPD, (b) usSDPD and (c) DPD.} Both SDPD and usSDPD simulations are performed at \SI{0.1}{\um} granularity, the former at time-scale \SI{1e-9}{s} and the latter at \SI{4e-8}{s}. The viscosity and the speed of sound are matched to those of water and the temperature is \SI{300}{K} in both cases. In the DPD simulation, the scales are chosen, such that the mapping number is $N_m = 10^6$, density is matched to the density of water and the temperature is \SI{300}{K}. The DPD parameter $a=10^7$ is obtained by matching the water speed of sound at the desired mapping number and $\gamma$ is set to 50. While both SDPD and usSDPD show a linear MSD, plain DPD freezes.}
    \label{fig:msd-curves}
\end{figure}

Figure~\ref{fig:msd-curves} shows that the mean-square displacement (MSD) curves exhibit linear behavior both for vanilla SDPD and usSDPD methods, indicating the absence of freezing artifacts at the studied scales. In contrast, ordinary DPD displays a horizontal MSD curve, signifying particle freezing. The diffusion coefficients derived here represent the effective diffusion of SDPD particles and do not directly correspond to physical diffusion constants. Nevertheless, as a first approximation, they can be interpreted using the Einstein relation for spherical particles in a low–Reynolds-number fluid,
\begin{equation}
	D = \frac{k_B T}{6 \pi \eta r},
\end{equation}
where $r$ denotes the particle radius. Using $T = \SI{300}{K}$, $r = \SI{2.5e-8}{m}$, and $\eta = \SI{8.9e-4}{\pascal \second}$ yields $D \approx \SI{1e-11}{m^2/s}$, differing by less than an order of magnitude among the two methods. For the usSDPD case, where $r = \SI{1.5e-8}{m}$, slightly higher diffusion coefficient is obtained, consistent with the results shown in Figure~\ref{fig:msd-icsdpd-lj-artpres}.

\begin{figure}[h!]
    \centering
    \begin{subfigure}[t]{0.49\textwidth}
        \centering
        \caption{}
        \includegraphics[width=\textwidth]{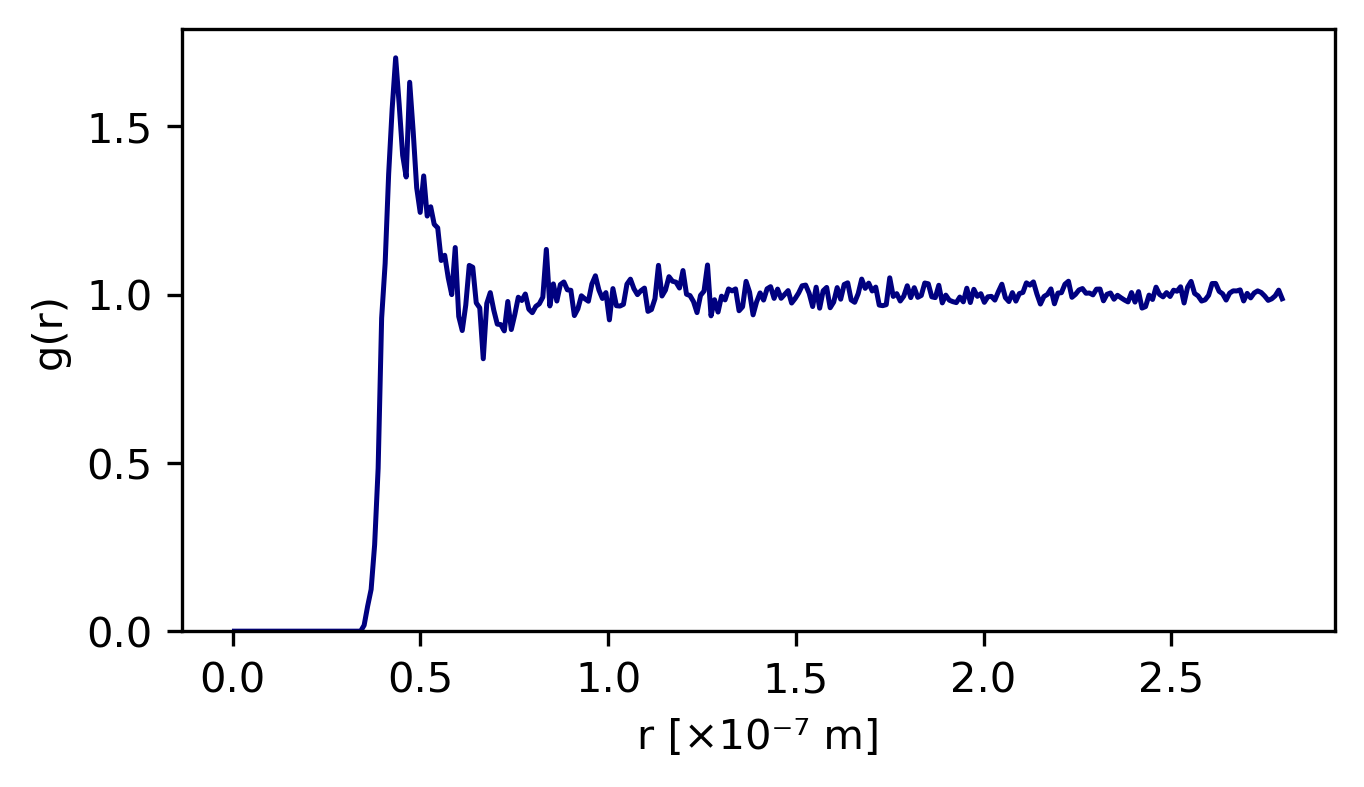}
        \label{fig:rdf-sdpd}
    \end{subfigure}
    \hfill
    \begin{subfigure}[t]{0.49\textwidth}
        \centering
        \caption{}
        \includegraphics[width=\textwidth]{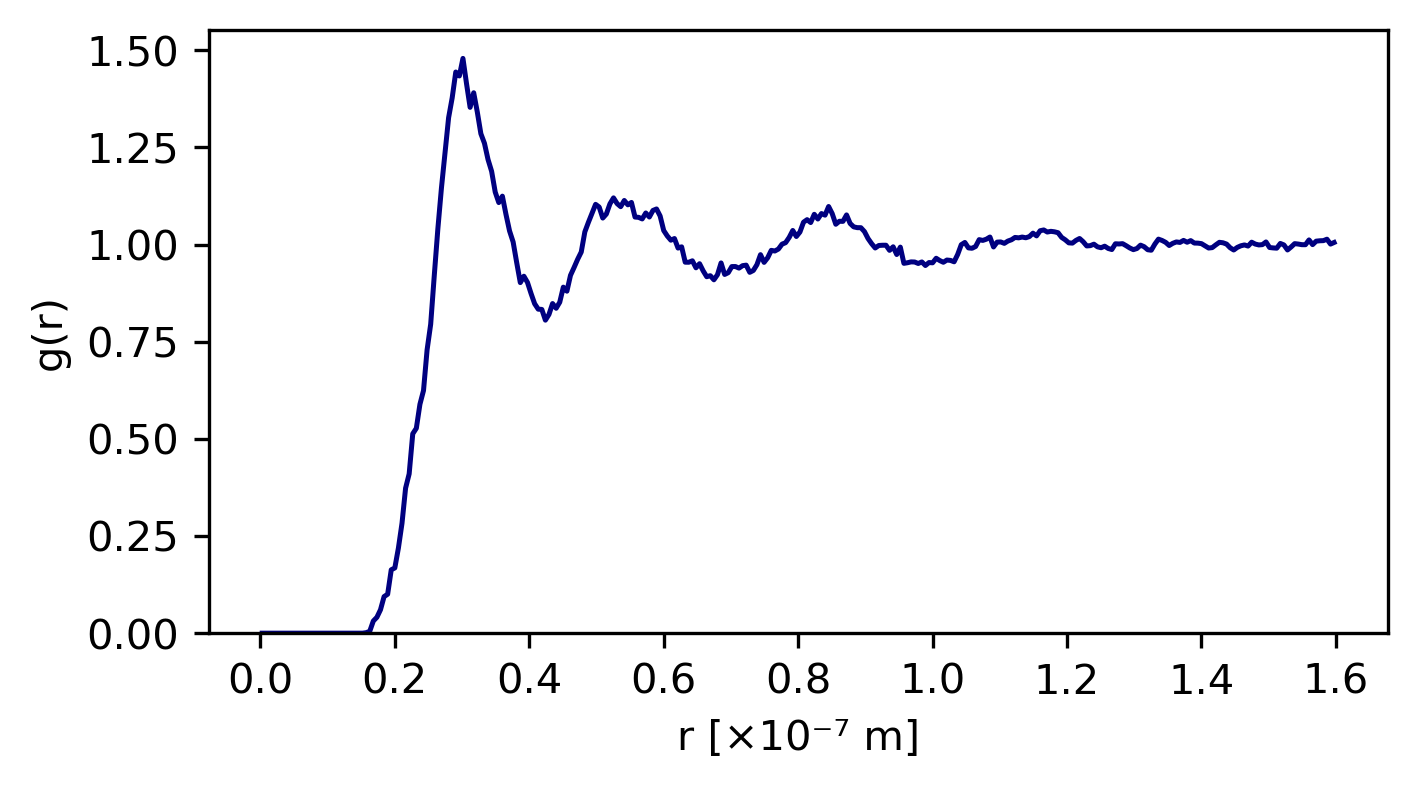}
        \label{fig:rdf-icsdpd-lj-artpres}
    \end{subfigure}

    \caption{\textbf{RDF curves for (a) vanilla SDPD and (b) usSDPD.}}
    \label{fig:rdf-curves}
\end{figure}

The radial distribution functions (RDFs) for the vanilla SDPD and usSDPD are shown in Figure~\ref{fig:rdf-curves}. In Figure~\ref{fig:rdf-sdpd}, the highest RDF peak is located at $r = \SI{0.5e-7}{m}$, while in Figure~\ref{fig:rdf-icsdpd-lj-artpres} the peak shifts to $r = \SI{0.3e-7}{m}$. These distances correspond to the diameters of the fluid particles used in each model.

To compare the stability of the vanilla SDPD and usSDPD methods at nanoscopic scales, we perform simulations at 10~nm and 1~nm granularities. These regimes are particularly relevant for studying the interaction of terahertz (THz) excitations with biological and soft-matter systems, including proteins, lipid membranes, and other nanoscale structures. As the length scale decreases, the amplitude of thermal fluctuations grows, requiring a reduction in the simulation timestep to maintain numerical stability. Table~\ref{tab:nano-stability} summarizes the maximum stable time scales for vanilla SDPD and usSDPD methods. At the 1~nm scale, the stability of the two methods rapidly deteriorates. Notably, while the methods maintain stability if timestep is sufficiently reduced, they offer no clear advantage over simpler alternatives, such as DPD. Given these observations, and in light of recent findings by Papež \textit{et al.}~\cite{Papez2023}, we conclude that SDPD-based methods are not well-suited for US simulations at nanometer scales. Instead, the standard DPD method provides a more efficient and robust alternative, and should be preferred in virtual US machine implementations operating at such resolutions.

\begin{table}[h!]
    \centering
    \begin{subtable}[t]{0.48\textwidth}
        \centering
        \caption{}
        \begin{tabular}{|l|c|p{5cm}|}
            \hline
            \textbf{Method} & \textbf{Max Time Scale} & \textbf{Comment} \\
            \hline
            vanilla SDPD & 0.5 ns & At timestep 1 ns, simulation crashes after some time. \\
            usSDPD       & 1 ns   &  \\
            \hline
        \end{tabular}
    \end{subtable}
    \hfill
    \begin{subtable}[t]{0.48\textwidth}
        \centering
        \caption{}
        \begin{tabular}{|l|c|p{5cm}|}
            \hline
            \textbf{Method} & \textbf{Max Time Scale} & \textbf{Comment} \\
            \hline
            vanilla SDPD & 0.001 ns & At timestep 0.01 ns, fluid freezes after some time. \\
            usSDPD       & 0.01 ns  &  \\
            \hline
        \end{tabular}
    \end{subtable}
    \caption{\textbf{Maximum stable time scale for the vanilla SDPD and usSDPD at nanoscopic length scales.} (a) 10 nm length scale, (b) 1 nm length scale. The speed of sound and viscosity match to those of water: $c = \SI{1481}{m/s}$ and $\eta = \SI{8.9e-4}{\pascal \second}$.}
    \label{tab:nano-stability}
\end{table}

\begin{figure}[h!]
    \centering
    \includegraphics[width=0.5\textwidth]{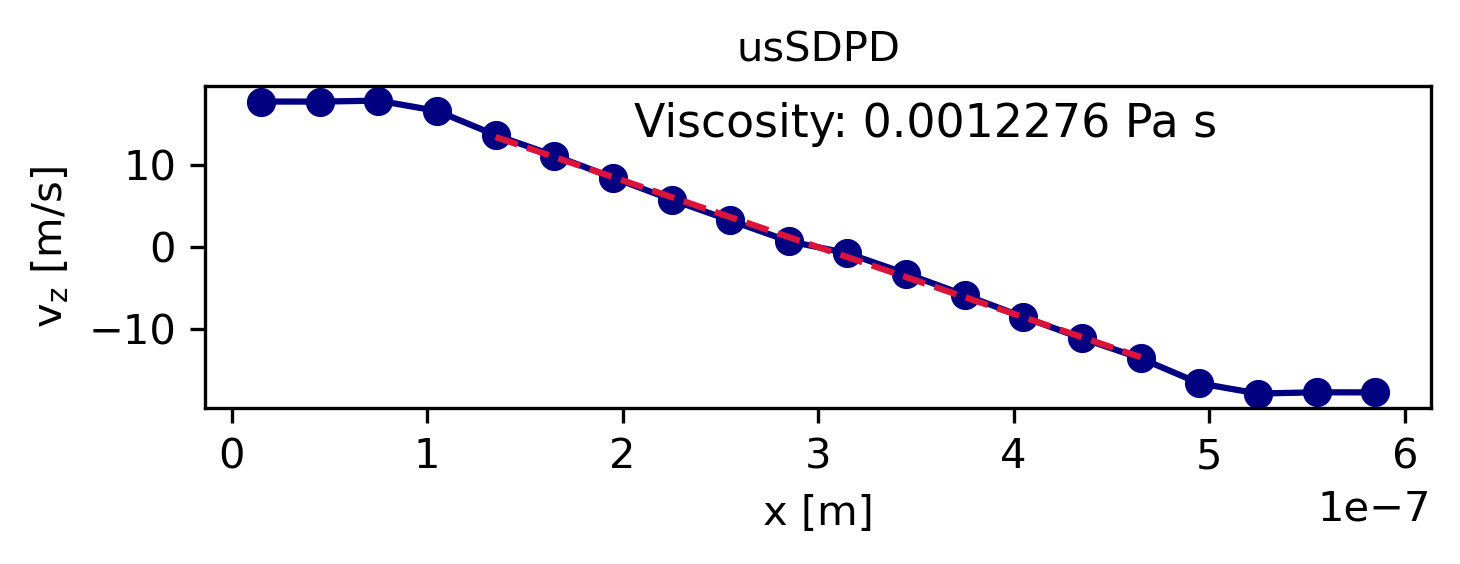}
    \caption{\textbf{Viscosity profile for usSDPD.} The plot includes the simulated viscosity, calculated from the slope of the profile.}
    \label{fig:viscosity-curves}
\end{figure}

To evaluate the accuracy of the measured viscosity relative to the input viscosity, we conduct a Couette flow simulation using the OBMD framework. Shear forces are applied within the buffer regions, and the resulting velocity profile is measured in the region of interest (see Figure \ref{fig:viscosity-snapshot}). Simulation parameters are listed in detail in the Supplementary Information, Section \ref{sec:parameters}. Figure \ref{fig:viscosity-curves} presents the velocity profile for the usSDPD method. As expected, the profile exhibits linear behavior, consistent with theoretical predictions. The simulated viscosity can be extracted from the slope of the profile and is annotated in the figure. We note that the simulated viscosity is slightly higher than the input viscosity, likely due to additional LJ interaction potential. For a slightly reduced input viscosity of \SI{6e-4}{\pascal \second}, the measured viscosity is \SI{7.1e-4}{\pascal \second} (not shown), which  demonstrates that, through careful tuning of the input viscosity, it is possible to achieve a simulated viscosity that closely approximates the actual viscosity of water.

\begin{figure}[h!]
    \centering
    \includegraphics[width=0.5\textwidth]{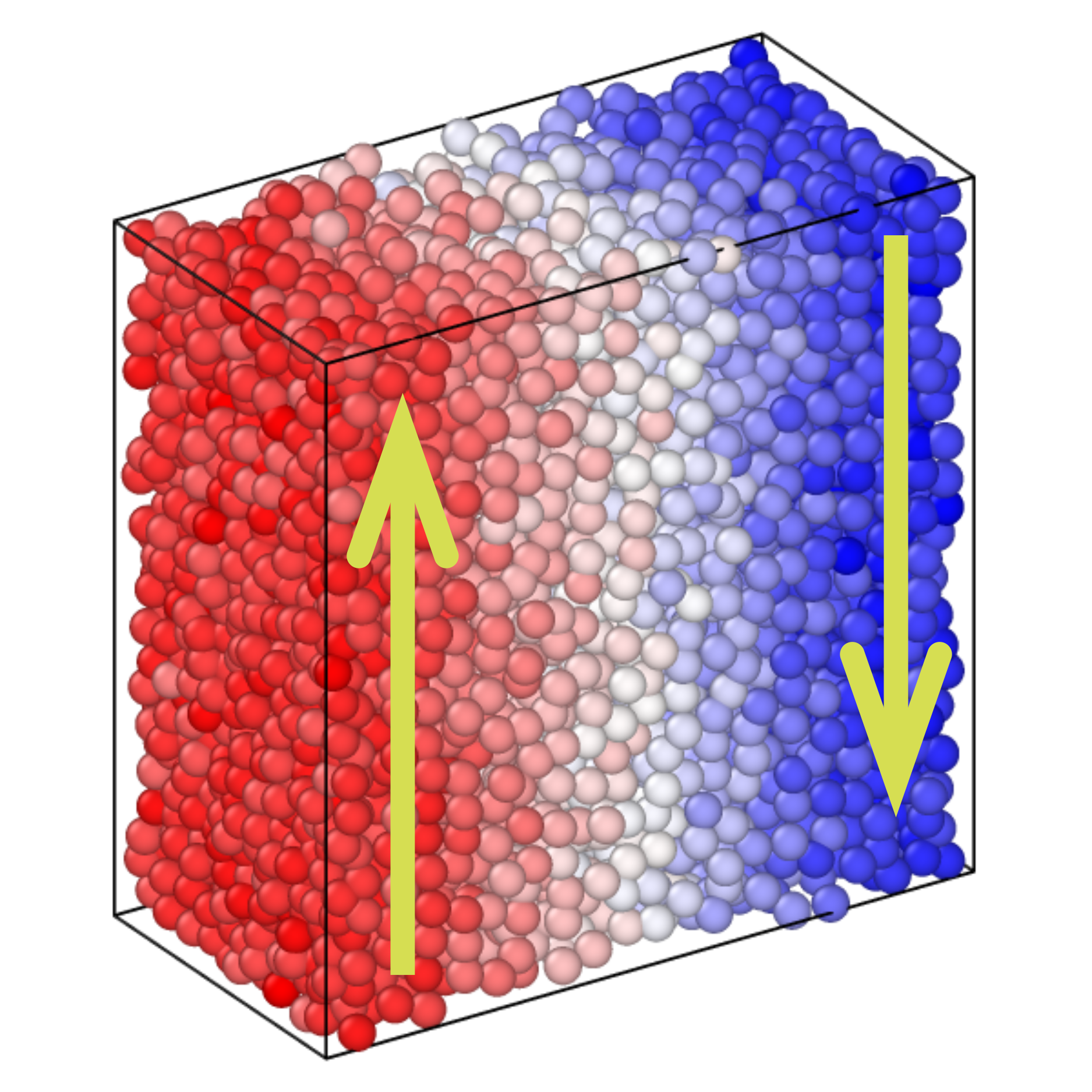}
    \caption{\textbf{Snapshot of the Couette flow simulation.} Particles are color-coded based on the vertical component of velocity (denoted with arrows).}
    \label{fig:viscosity-snapshot}
\end{figure}

\begin{figure}[h!]
    \begin{subfigure}[t]{0.5\textwidth}
        \centering
        \caption{}
        \includegraphics[width=\textwidth]{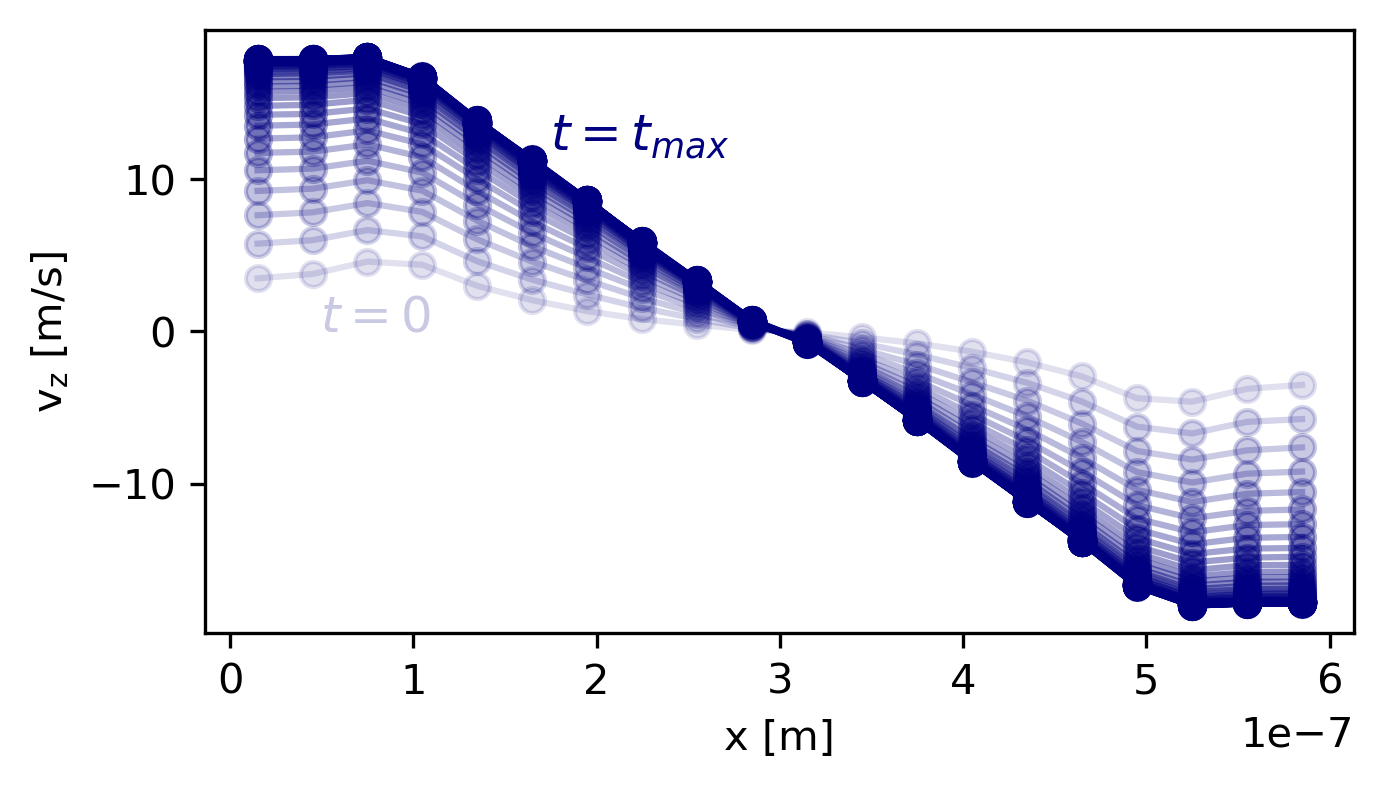}
        \label{fig:viscosity-equilibration1}
    \end{subfigure}
	\hfill    
    \begin{subfigure}[t]{0.5\textwidth}
        \centering
        \caption{}
        \includegraphics[width=\textwidth]{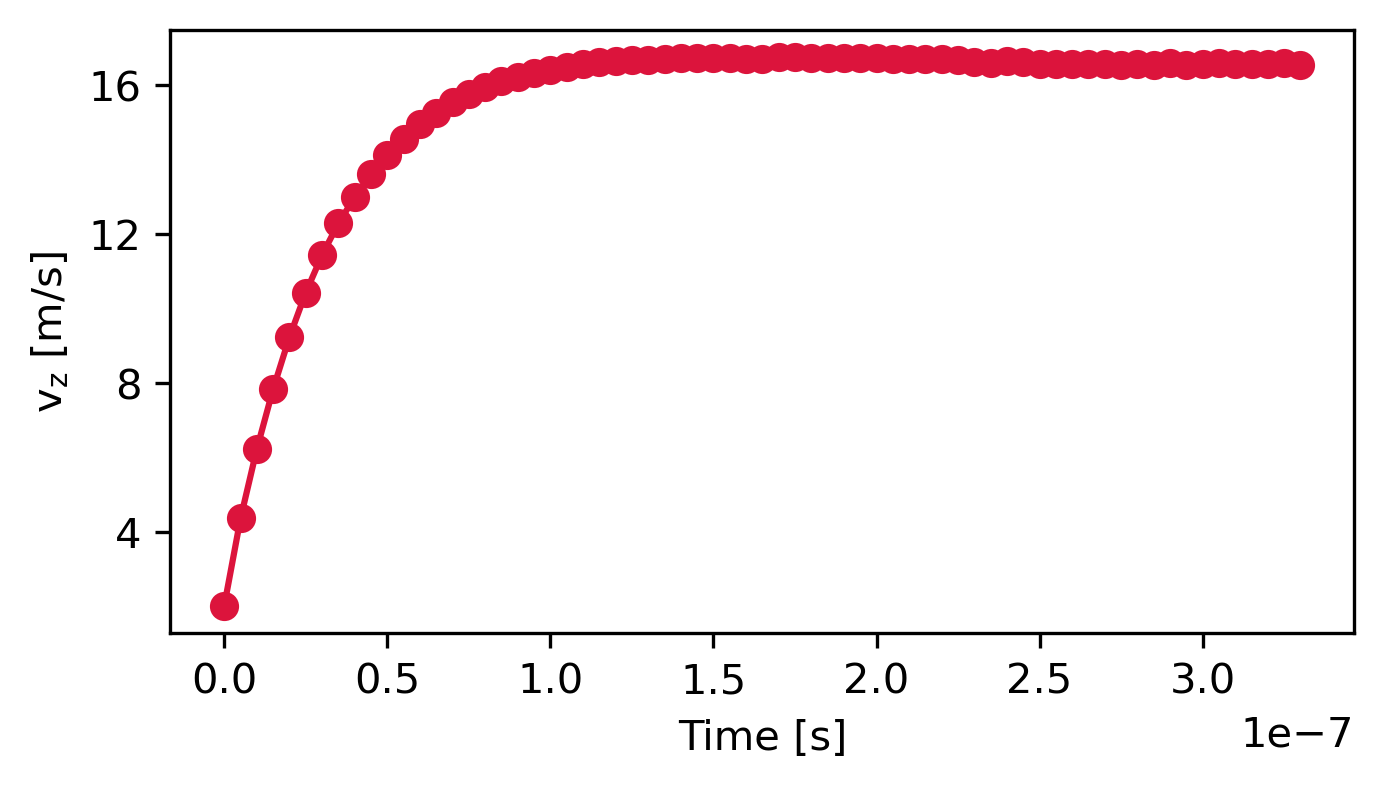}
        \label{fig:viscosity-equilibration2}
    \end{subfigure}
    
    \caption{\textbf{Viscosity equilibration for usSDPD.} (a) Evolution of velocity profile. (b) Temporal evolution of velocity at a fixed position $x$.}
    \label{fig:viscosity-equilibration}
\end{figure}

Figure~\ref{fig:viscosity-equilibration} illustrates the establishment of the linear velocity profile in the Couette flow experiment. The profile forms after approximately \SI{1e-7}{s}, consistent with the analytical estimate of the characteristic start-up time $\tau = \rho h^2 / \eta$, where $h$ is the gap between the plates (corresponding to the simulation box width). For a width of \SI{0.6}{\um} and shear buffer thickness of \SI{0.075}{\um}, we obtain $\tau \approx \SI{1.7e-7}{s}$, which is in agreement with the simulation results.

\begin{figure}[h!]
    \centering
    \includegraphics[width=0.3\textwidth]{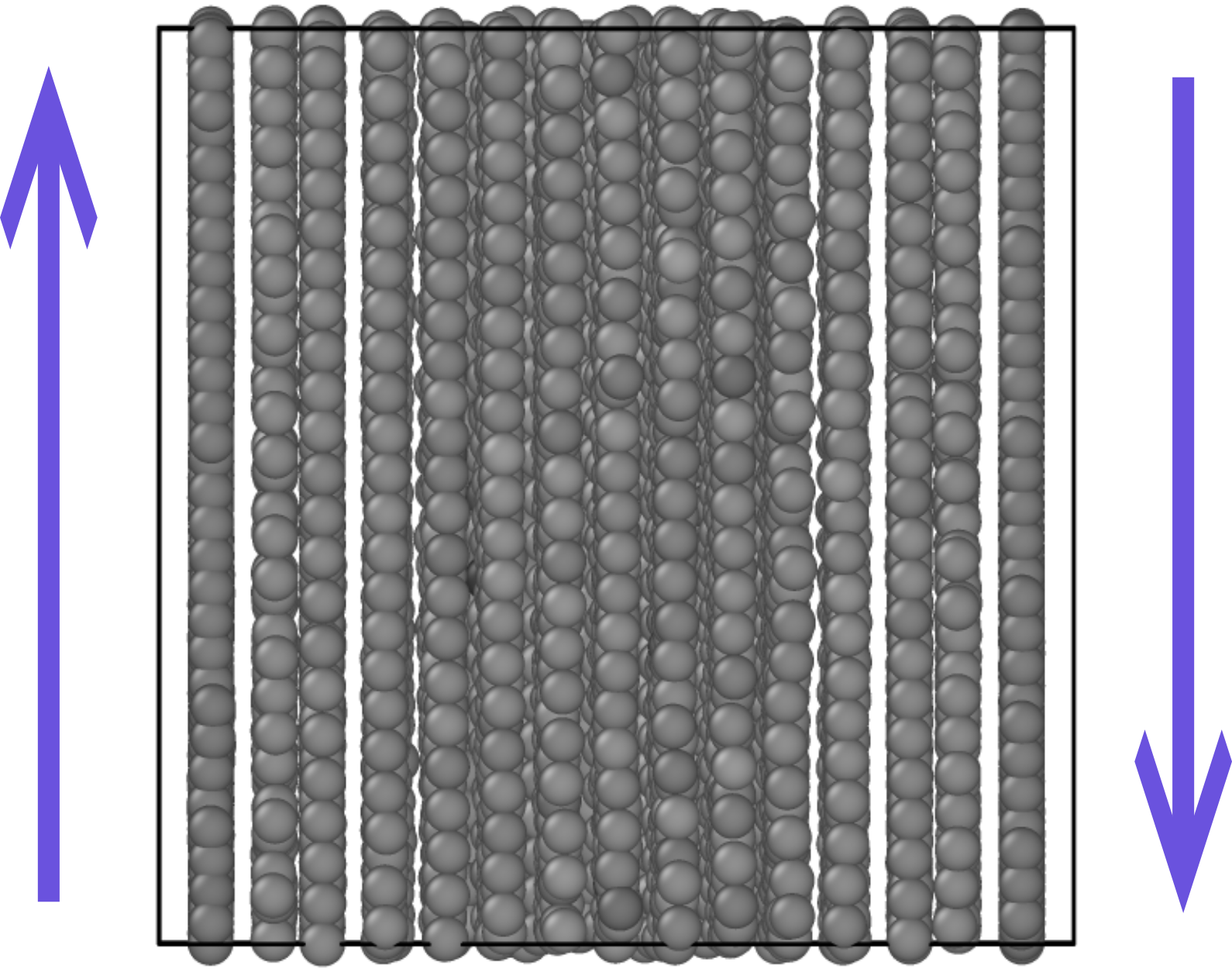}
    \caption{\textbf{Snapshot of equilibrium Couette flow at \SI{1}{\um} granularity.} Arrows denote the direction of the applied shear force.}
    \label{fig:viscosity-icsdpd-lj-artpres-largescale-snapshot}
\end{figure}

At the \SI{1}{\um} granularity, the measured viscosity of the usSDPD fluid agrees well with the values obtained at the \SI{0.1}{\um} granularity, however, the reduced thermal fluctuations lead to the formation of layered structures within the Couette flow (Figure~\ref{fig:viscosity-icsdpd-lj-artpres-largescale-snapshot}), a phenomenon not observed at smaller scales where fluctuations promote particle mixing along the shear velocity gradient. The theoretically predicted equilibration time at this scale is more than five times longer than the value obtained in simulation. Consequently, simulations at the \SI{0.1}{\um} granularity are employed throughout this study, as they most faithfully reproduce the physical properties of water.

\subsection{Parameters}
\label{sec:parameters}

This subsection summarizes the parameters used in the simulations presented in this paper. Tables \ref{tab:parameters-stability-analysis} and \ref{tab:parameters-viscosity-analysis} list the parameters employed in the stability and viscosity analyses described in Section \ref{sec:fluid-simulation}, respectively. The parameters used in the acoustophoresis simulations are given in Table \ref{tab:parameters-acoustophoresis}.

\begin{table}[htbp]
    \centering
    \caption{Common simulation parameters for stability analysis simulations. Parameters beginning with a capital case letter are given in SI units, parameters beginning with a lower case letter are given in dimensionless units. Values in black color are common to both methods, values in green are specific to SDPD and values in blue are specific to usSDPD. Some of the technical parameters are skipped for brevity.}
    \renewcommand{\arraystretch}{1.2}
    \begin{tabular}{>{\raggedright\arraybackslash}p{3cm} >{\raggedright\arraybackslash}p{2.5cm} >{\raggedright\arraybackslash}p{6cm}}
        \hline
        \textbf{Parameter} & \textbf{Value} & \textbf{Short description} \\
        \hline
        \multicolumn{3}{l}{\textbf{Length scale}} \\
        \texttt{Length\_scale} & $1.0\times10^{-7}$ & kernel function cutoff $h$ [m] \\
        \hline
        \multicolumn{3}{l}{\textbf{Outer fluid parameters}} \\
        \texttt{Fluid\_density} & 998.0 & density of water [kg/m$^3$] \\
        \texttt{Viscosity} & $8.9\times10^{-4}$ & dynamic viscosity of water [Pa s] \\
        \texttt{Speed\_of\_sound} & 1481.0 & speed of sound [m/s] \\
        \texttt{Temperature} & 300.0 & temperature [K] \\
        \texttt{Kb} & $1.3806504\times10^{-23}$ & Boltzmann constant [J/K] \\
        \hline
        \multicolumn{3}{l}{\textbf{Simulation parameters}} \\
        \texttt{skin} & \textcolor{green}{0.2} or \textcolor{blue}{0.12} & skin thickness \\
        \texttt{particle\_diameter} & \textcolor{green}{0.5} or \textcolor{blue}{0.3} & particle diameter \\
        \texttt{timestep} & 0.001 & timestep size relative to the time scale \\
        \texttt{max\_rho\_relative\_error} & \textcolor{blue}{0.1} & terminating condition for pressure iterations (max cumulative density error $<$ 0.1\%) \\
        \texttt{omega} & \textcolor{blue}{0.5} & relaxation parameter for Jacobi pressure iterations \\
        \texttt{psi} & \textcolor{blue}{1.5} & parameter of the implicit compressible pressure solver \\
        \hline
        \multicolumn{3}{l}{\textbf{Lennard-Jones potential that prevents pair instability}} \\
        \texttt{lj\_cutoff} & \textcolor{blue}{0.6} & cutoff distance \\
        \texttt{lj\_epsilon} & \textcolor{blue}{$1.0\times10^{-5}$} & depth of the potential well \\
        \texttt{lj\_sigma} & \textcolor{blue}{0.23} & finite distance at which the potential is zero \\
        \hline
    \end{tabular}
    \label{tab:parameters-stability-analysis}
\end{table}

\begin{table}[htbp]
    \centering
    \caption{Simulation parameters for viscosity analysis simulation. Parameters beginning with a capital case letter are given in SI units, parameters beginning with a lower case letter are given in dimensionless units. Some of the technical parameters are skipped for brevity.}
    \renewcommand{\arraystretch}{1.2}
    \begin{tabular}{>{\raggedright\arraybackslash}p{3cm} >{\raggedright\arraybackslash}p{2.5cm} >{\raggedright\arraybackslash}p{6cm}}
        \hline
        \textbf{Parameter} & \textbf{Value} & \textbf{Short description} \\
        \hline
        \multicolumn{3}{l}{\textbf{Length scale}} \\
        \texttt{Length\_scale} & $1.0\times10^{-7}$ & kernel function cutoff $h$ [m] \\
        \hline
        \multicolumn{3}{l}{\textbf{Outer fluid parameters}} \\
        \texttt{Fluid\_density} & 998.0 & density of water [kg/m$^3$] \\
        \texttt{Viscosity} & $8.9\times10^{-4}$ & dynamic viscosity of water [Pa s] \\
        \texttt{Speed\_of\_sound} & 1481.0 & speed of sound [m/s] \\
        \texttt{Temperature} & 300.0 & temperature [K] \\
        \texttt{Kb} & $1.3806504\times10^{-23}$ & Boltzmann constant [J/K] \\
        \hline
        \multicolumn{3}{l}{\textbf{Simulation parameters}} \\
        \texttt{skin} & 0.12 & skin thickness \\
        \texttt{particle\_diameter} & 0.3 & particle diameter \\
        \texttt{timestep} & 0.001 & timestep size relative to the time scale \\
        \texttt{max\_rho\_relative\_error} & 0.1 & terminating condition for pressure iterations (max cumulative density error $<$ 0.1\%) \\
        \texttt{omega} & 0.5 & relaxation parameter for Jacobi pressure iterations \\
        \texttt{psi} & 1.5 & parameter of the implicit compressible pressure solver \\
        \hline
        \multicolumn{3}{l}{\textbf{Lennard-Jones potential that prevents pair instability}} \\
        \texttt{lj\_cutoff} & 0.6 & cutoff distance \\
        \texttt{lj\_epsilon} & $1.0\times10^{-5}$ & depth of the potential well \\
        \texttt{lj\_sigma} & 0.23 & finite distance at which the potential is zero \\
        \hline
        \multicolumn{3}{l}{\textbf{OBMD parameters}} \\
        \texttt{pxx} & 1.0 & background pressure \\
        \texttt{pxy} & 0.0 & external xy shear stress \\
        \texttt{pxz} & 1.0 & external xz shear stress \\
        \texttt{buffer\_relative\_size} & 0.2 & buffer size relative to the simulation cell length \\
        \texttt{shear\_relative\_size} & 0.125 & shear size relative to the simulation cell length \\
        \texttt{alpha} & 0.7 & desired number of particles in the buffer scaled by \texttt{alpha} \\
        \texttt{tau\_usher} & 30 & relaxation time of the insertion algorithm \\
        \hline
    \end{tabular}
    \label{tab:parameters-viscosity-analysis}
\end{table}

\begin{table}[htbp]
    \centering
    \caption{Common simulation parameters for all acoustophoresis simulations. Parameters beginning with a capital case letter are given in SI units, parameters beginning with a lower case letter are given in dimensionless units. Some of the technical parameters are skipped for brevity.}
    \renewcommand{\arraystretch}{1.2}
    \begin{tabular}{>{\raggedright\arraybackslash}p{3cm} >{\raggedright\arraybackslash}p{2.5cm} >{\raggedright\arraybackslash}p{6cm}}
        \hline
        \textbf{Parameter} & \textbf{Value} & \textbf{Short description} \\
        \hline
        \multicolumn{3}{l}{\textbf{Length and time scales}} \\
        \texttt{Length\_scale} & $1.0\times10^{-7}$ & kernel function cutoff $h$ [m] \\
        \texttt{Time\_scale} & $1.0\times10^{-8}$ & time scale [s] \\
        \hline
        \multicolumn{3}{l}{\textbf{Outer fluid parameters}} \\
        \texttt{Fluid\_density} & 998.0 & density of water [kg/m$^3$] \\
        \texttt{Speed\_of\_sound} & 1481.0 & speed of sound [m/s] \\
        \texttt{Temperature} & 300.0 & temperature [K] \\
        \texttt{Kb} & $1.3806504\times10^{-23}$ & Boltzmann constant [J/K] \\
        \hline
        \multicolumn{3}{l}{\textbf{Simulation parameters}} \\
        \texttt{skin} & 0.12 & skin thickness \\
        \texttt{particle\_diameter} & 0.3 & particle diameter \\
        \texttt{timestep} & 0.003 & timestep size relative to the time scale \\
        \texttt{max\_rho\_relative\_error} & 0.1 & terminating condition for pressure iterations (max cumulative density error $<$ 0.1\%) \\
        \texttt{omega} & 0.5 & relaxation parameter for Jacobi pressure iterations \\
        \texttt{psi} & 1.5 & parameter of the implicit compressible pressure solver \\
        \hline
        \multicolumn{3}{l}{\textbf{Lennard-Jones potential that prevents pair instability}} \\
        \texttt{lj\_cutoff} & 0.6 & cutoff distance \\
        \texttt{lj\_epsilon} & $1.0\times10^{-5}$ & depth of the potential well \\
        \texttt{lj\_sigma} & 0.23 & finite distance at which the potential is zero \\
        \hline
        \multicolumn{3}{l}{\textbf{DPD parameters (gas inside microbubble) for slow internal bubble dynamics}} \\
        \texttt{dpd\_temperature} & 0.001 & DPD temperature \\
        \texttt{dpd\_cutoff} & 0.6 & DPD cutoff distance \\
        \texttt{dpd\_a\_water\_gas} & 2.0 & conservative force parameter between water and gas \\
        \texttt{dpd\_gamma\_water\_gas} & 10.0 & dissipative force parameter between water and gas \\
        \texttt{dpd\_a\_gas\_gas} & 2.0 & conservative force parameter between gas particles \\
        \texttt{dpd\_gamma\_gas\_gas} & 10.0 & dissipative force parameter between gas particles \\
        \hline
        \multicolumn{3}{l}{\textbf{Lennard-Jones parameters for gas-shell interaction}} \\
        \texttt{lj\_cutoff2} & 0.6 & cutoff distance \\
        \texttt{lj\_epsilon2} & 0.01 & depth of the potential well \\
        \texttt{lj\_sigma2} & 0.3 & finite distance at which the potential is zero \\
        \hline
        \multicolumn{3}{l}{\textbf{OBMD parameters}} \\
        \texttt{buffer\_relative\_size} & 0.1 & buffer size relative to the simulation cell length \\
        \texttt{shear\_relative\_size} & 0.125 & shear size relative to the simulation cell length \\
        \texttt{alpha} & 0.7 & desired number of particles in the buffer scaled by \texttt{alpha} \\
        \texttt{tau\_usher} & 30 & relaxation time of the insertion algorithm \\
        \hline
        \multicolumn{3}{l}{\textbf{Microbubble material}} \\
        \texttt{Shell\_density} & 1100.0 & density of the shell material [kg/m$^3$] \\
        \texttt{Shell\_thickness} & $15.0\times10^{-9}$ & thickness of the shell [m] \\
        \texttt{viscosity\_shell} & 0.1 & small additional viscous force on shell particles \\
        \hline
        \multicolumn{3}{l}{\textbf{Microbubble elasticity}} \\
        \texttt{Yt} & $2.64\times10^{7}$ & Young's modulus \\
        \texttt{nu} & 0.5 & Poisson ratio \\
        \texttt{fscale} & 0.0003 & scaling factor in isotropic elastic model \\
        \hline
        \multicolumn{3}{l}{\textbf{Microbubble inner fluid}} \\
        \texttt{Gas\_density} & 998.0 & density of the inner fluid \\
        \hline
    \end{tabular}
    \label{tab:parameters-acoustophoresis}
\end{table}

\end{document}